\documentclass[11pt,preprint]{aastex}
\usepackage{natbib}
\usepackage{epsfig}
\makeatletter
\renewcommand\@biblabel[1]{\hspace{-\labelsep}}
\makeatother

\def\etal{{\it et al.}}
\def\deg{$^\circ$}

\begin{document}


\title{~~\\ ~~\\ First Light for the First Station of the Long Wavelength Array }
\shorttitle{First Light for LWA1}
\shortauthors{Taylor et al.}
\author{G.B. Taylor\altaffilmark{1,2},  S. W. Ellingson\altaffilmark{3},
N.E. Kassim\altaffilmark{4}, J. Craig\altaffilmark{1}, 
J. Dowell\altaffilmark{1}, 
C. N. Wolfe\altaffilmark{3},
J. Hartman\altaffilmark{5,6},
G. Bernardi\altaffilmark{7},
T. Clarke\altaffilmark{4},
A. Cohen\altaffilmark{8},
N.P. Dalal\altaffilmark{9},
W.C. Erickson\altaffilmark{10},
B. Hicks\altaffilmark{4},
L.J. Greenhill\altaffilmark{7},
B. Jacoby\altaffilmark{11},
W. Lane\altaffilmark{4},
J. Lazio\altaffilmark{5},
D. Mitchell\altaffilmark{12},
R. Navarro\altaffilmark{5},
S. M. Ord\altaffilmark{7},
Y. Pihlstr\"om\altaffilmark{1}, 
E. Polisensky\altaffilmark{4},
P.S. Ray\altaffilmark{4},
L.J. Rickard\altaffilmark{1},   
F.K. Schinzel\altaffilmark{1},
H. Schmitt\altaffilmark{13},
E. Sigman\altaffilmark{5},   
M. Soriano\altaffilmark{5},
K.P. Stewart\altaffilmark{4},
K. Stovall\altaffilmark{14},
S. Tremblay\altaffilmark{15,16},
D. Wang\altaffilmark{5},
K.W. Weiler\altaffilmark{4},
S. White\altaffilmark{17},
D.L. Wood\altaffilmark{18}
}

\affil{}
\email{contact author: gbtaylor@unm.edu}
\altaffiltext{1}{Department of Physics and Astronomy, University of New
Mexico, Albuquerque NM, 87131, USA}
\altaffiltext{2}{Greg Taylor is also an Adjunct Astronomer at the National Radio Astronomy Observatory.}
\altaffiltext{3}{Bradley Department of Electrical and Computer Engineering, Virginia Polytechnic Institute and State University, Blacksburg, VA 24061, USA}
\altaffiltext{4}{U.S. Naval Research Laboratory, Washington, DC 20375 USA}
\altaffiltext{5}{NASA Jet Propulsion Laboratory, Pasadena, CA 91109 USA}
\altaffiltext{6}{NASA Postdoctoral Program fellow}
\altaffiltext{7}{Harvard-Smithsonian Center for Astrophysics, 60 Garden St., Cambridge, MA 02138, USA}
\altaffiltext{8}{Johns Hopkins University Applied Physics Laboratory
Laurel, MD 20723 USA}
\altaffiltext{9}{Northrup Grumman, Aerospace Systems, Redondo Beach, CA 90278 USA}
\altaffiltext{10}{University of Tasmania, Hobart, Australia}
\altaffiltext{11}{Affiliated with The Aerospace Corporation, Chantilly, VA 20151 USA}
\altaffiltext{12}{University of Melbourne, Australia}
\altaffiltext{13}{Computational Physics, Inc., Springfield, VA 22151 USA}
\altaffiltext{14}{Center for Gravitational Wave Astronomy and Department of Physics and Astronomy, University of Texas at Brownsville, Brownsville, TX 78520, USA}
\altaffiltext{15}{ARC Centre of Excellence for All-sky Astrophysics (CAASTRO)}
\altaffiltext{16}{International Centre for Radio Astronomy Research, 
Curtin University, Bentley WA, Australia}
\altaffiltext{17}{Space Vehicles Directorate, AFRL, Albuquerque, NM USA}
\altaffiltext{18}{Praxis, Inc., Alexandria, VA 22303 USA}


\begin{abstract}
  The first station of the Long Wavelength Array (LWA1) was completed
  in April 2011 and is currently performing observations resulting
  from its first call for proposals in addition to a continuing
  program of commissioning and characterization observations.  The
  instrument consists of 258 dual-polarization dipoles, which are
  digitized and combined into beams.  Four independently-steerable
  dual-polarization beams are available, each with two “tunings” of 16
  MHz bandwidth that can be independently tuned to any frequency
  between 10 MHz and 88 MHz.  The system equivalent flux density for
  zenith pointing is $\sim$3 kJy and is approximately independent of
  frequency; this corresponds to a sensitivity of $\sim$5 Jy/beam
  (5$\sigma$, 1 s); making it one of the most sensitive
  meter-wavelength radio telescopes.  LWA1 also has 
two ``transient buffer'' modes
  which allow coherent recording from all dipoles simultaneously,
  providing instantaneous all-sky field of view.  LWA1 provides versatile
  and unique new capabilities for Galactic science, pulsar science,
  solar and planetary science, space weather, cosmology, and searches
  for astrophysical transients.  Results from LWA1 will detect or
  tightly constrain the presence of hot Jupiters within 50 parsecs of
  Earth.  LWA1 will provide excellent resolution in frequency
  and in time to examine phenomena such as solar bursts, 
 and pulsars over a 4:1 frequency range that
  includes the poorly understood turnover and steep-spectrum regimes.
  Observations to date have proven LWA1's potential for pulsar
  observing, and just a few seconds with the completed 256-dipole LWA1
  provide the most sensitive images of the sky at 23 MHz obtained
  yet. We are operating LWA1 as an {\it open skies} radio observatory,
  offering $\sim$2000 beam-hours per year to the general community.  
  At the same time, we are
  operating a backend for all-sky imaging and total-power
  transient detection, approximately 6840 hours per year ($\sim$78\% duty
  cycle).
\end{abstract}

\section{Introduction to LWA1}

LWA1 originated as the first ``station'' (beamforming array) of the
Long Wavelength Array (LWA).  The LWA concept was conceived by 
Perley \& Erickson (1984) and expanded by Kassim \& Erickson (1998) 
and Kassim \etal\ (2005).  It gained momentum with sub-arcminute
imaging with the VLA at 74 MHz (Kassim \etal\ 1993, 2007a) and 
the project began in earnest in April 2007,
sponsored primarily by the Office of Naval Research (ONR), with the
ultimate goal of building an aperture synthesis radio telescope
consisting of 53 identical stations distributed over the
U.S. Southwest (Ellingson \etal\ 2009, 2011).  
Currently, the
LWA project exists as a collaboration, which we refer
to as the Long Wavelength Array Collaboration (LWAC).  The LWAC is an
informal ``umbrella'' organization which serves to facilitate
collaboration among organizations, projects, and individuals with
LWA-related interests.  The LWAC has no legal identity, no specific
charter, and no dedicated funding; membership is completely open and
it exists solely to facilitate collaboration among its members.  The
LWAC coordinator is currently N.E. Kassim of the U.S. Naval Research
Laboratory. The LWAC currently includes the following projects:
\begin{itemize}
\item LWA1: Originally conceived as the first station of the LWA; but which is a now a dedicated radio telescope distinct from (but supportive of) the separate effort to build a long-baseline aperture synthesis instrument.  The topic of this paper.
\item LoFASM; an initiative by the University of Texas at Brownsville (UTB), PI R. Jenet, to build and deploy three or more ``mini-stations'' for the purpose of monitoring the sky for transient sources.  
\item Cosmic Dawn; a project to measure, or place constraints, on the HI signal from the Dark Ages using beamforming data from LWA1.  This effort is being led by J. Bowman.
\item LEDA (``Large aperture Experiment to detect the Dark Ages''); an initiative to develop an alternative backend for LWA1 led by L. Greenhill.  Additional information about LEDA is provided in \S~2.4.
\item LWA; the continuing initiative to fund and build a 53-station long-baseline aperture synthesis instrument.
\end{itemize}

Institutions represented in the LWAC (as determined by attendance at
the May 12, 2011 LWA1 User Meeting) include U.S. Air Force Research
Laboratory (AFRL), Arizona State University (ASU), Harvard University, 
Kansas University (KU), Long Island
University, National Radio Astronomy Observatory (NRAO), NASA Jet
Propulsion Laboratory (JPL), U.S. Naval Research Laboratory (NRL), New Mexico
Tech (NMT), University of New Mexico (UNM), University of Texas
at Brownsville (UTB), and Virginia Tech (VT).
New institutions and individuals are invited to join the LWAC and 
if interested should contact Namir Kassim (NRL) or Greg Taylor (UNM).

\begin{figure}[tp]
\begin{center}
\includegraphics[width=5.9in,angle=0]{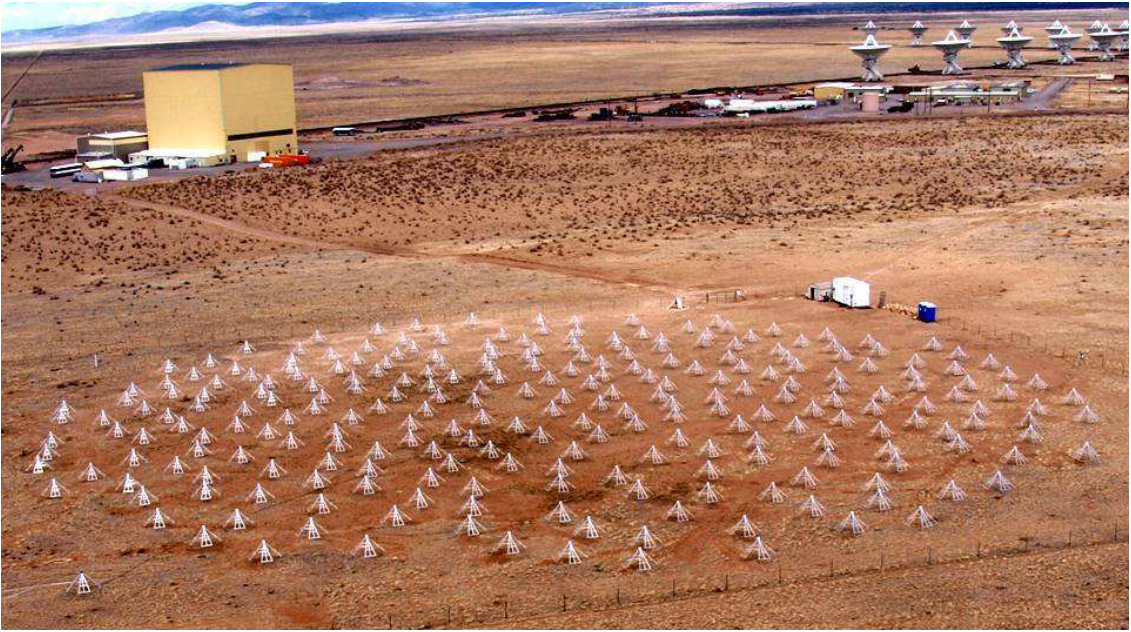}
\end{center}
\vspace{-0.5cm}
  \caption{
Aerial view of the LWA1 radio observatory. 
}
\label{fig:lwa_design}
\end{figure}

The LWA1 Radio Observatory is shown in Fig.~\ref{fig:lwa_design}.  It
is located on NRAO land within the central square mile of the VLA,
which offers numerous advantages.  The project to design and build
LWA1 was led by UNM, who also developed analog receivers and the
shelter and site infrastructure systems.  The system architecture was
developed by VT, who also developed LWA1's monitor \& control and data
recording systems.  Key elements of LWA1's design were guided by
experience gained from a prototype stub system project known as 
the Long Wavelength Demonstrator Array,
developed by NRL and the University of Texas at Austin (York \etal\
2007); and by VT's Eight-meter wavelength Transient Array (ETA;
Deshpande 2009).  NRL developed LWA1's active antennas, and JPL
developed LWA1's digital processing subsystem.  
LWA1 has recently been established as a University Radio Observatory
by NSF and as such will make regular calls for proposals from the 
astronomical community starting in February 2012.
Table~1 summarizes the capabilities of LWA.  For more details see
the LWA web pages at {\tt http://lwa.unm.edu} including the LWA
Memo series.


\begin{figure}[t!]
\begin{center}
\includegraphics[width=6.0in,angle=0]{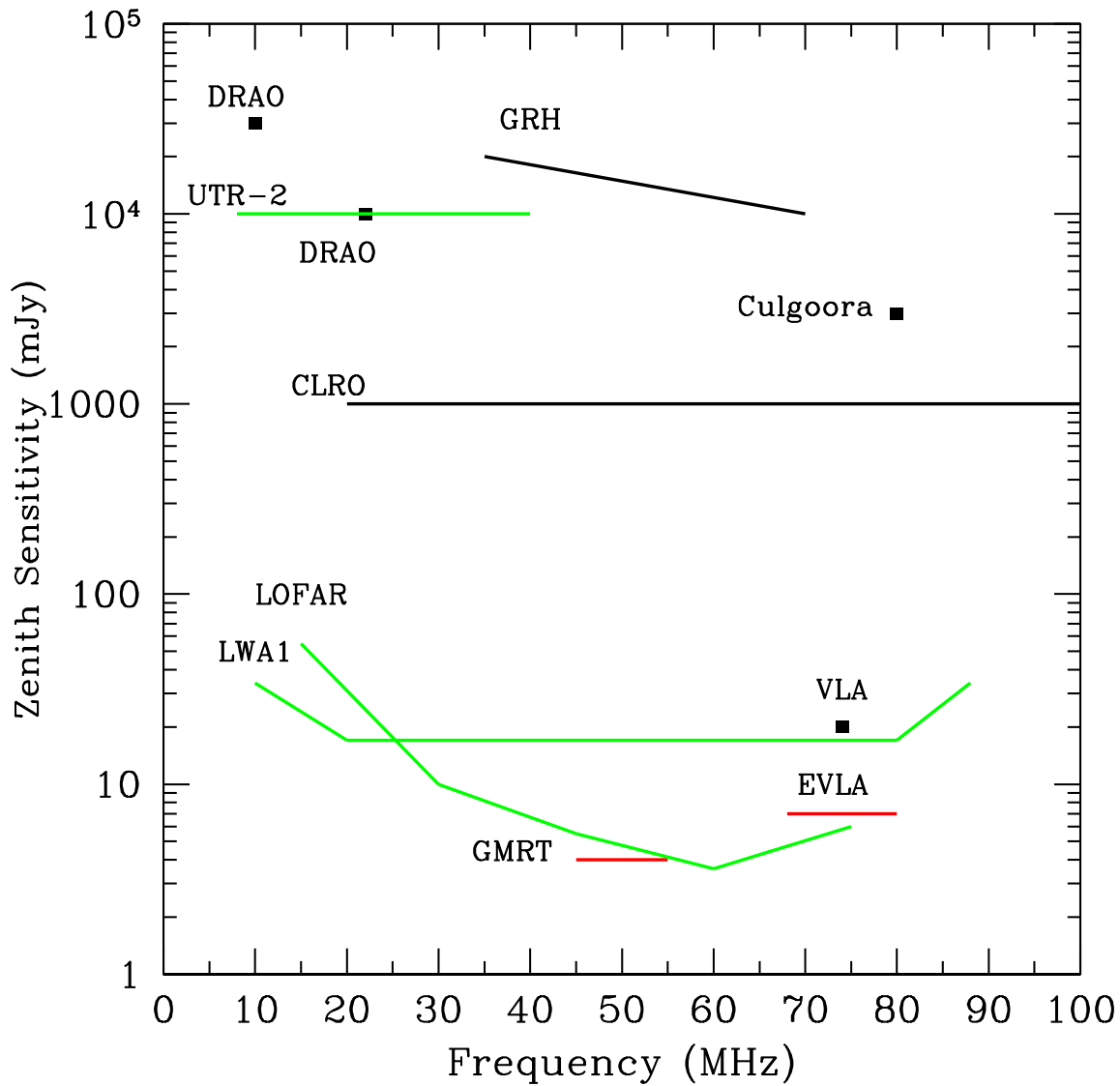}
\end{center}
\vspace{-0.5cm}
  \caption{
Sensitivity of the LWA1 compared to other instruments.  Integration time
is 1 hour for all instruments and the bandwidths assumed for current and 
planned instruments (green and red respectively) are as follows:
UTR2: $\sim$3 MHz, LOFAR: 16 MHz, Y=VLA: 3 MHz, LWA1: 16 MHz, GMRT: 10 MHz.
No effects of confusion noise are considered.
}
\label{fig:sens}
\end{figure}

Given the architectural similarities to a LOFAR low band array (LBA)
station (de Vos, Gunst \& Nijboer 2009), it is appropriate to identify
key similarities and differences.  LOFAR is an operational radio
telescope array based in the Netherlands 
with frequency coverage overlapping that of
LWA1. “AART-FAAC” (www.aartfaac.org) is a planned imaging-based
search for radio transients with LOFAR, using a correlator which is
currently under development. An approximate comparison of the
sensitivity of LOFAR and LWA1, based on information provided in
Wijnholds \& van Cappellen (2011), proceeds as follows: LWA1 has 258
dual-polarized dipoles (or ``stands''), whereas LOFAR currently has 36 stations consisting of 96
dual-polarized dipoles each, of which only 48 can be used
simultaneously, for a total of 1728 dual-polarized dipoles. However,
LWA1 per-dipole system temperature is dominated by Galactic noise by a
factor of at least 4:1, whereas LOFAR per-dipole system temperature is
at best 1:1. Taking this into account, we estimate all of LOFAR is
more sensitive than LWA1 by a factor of no greater than $\sim$4 on a
per-bandwidth basis, at LOFAR’s optimum frequency ($\sim$60 MHz). 
We also note that LWA1 (34\deg N) sees a significantly
different portion of the sky than LOFAR ($\sim$52\deg N), including access to
the interesting Galactic center and inner plane regions.
A graphical comparison of LWA1 with LOFAR and
other contemporaneous instruments is shown in Fig.~\ref{fig:sens}.

\begin{table}[t!]
\begin{center}
\small
\caption{Summary of LWA1 Specifications}
\vspace{0.1cm}
\begin{tabular}{ll}
Specification & As Built Description \\
\hline 
\hline
\vspace{-0.2cm}\\
Beams: & 4, independently-steerable, each dual-polarization \\ 
Tunings: & 2 independent center frequencies per beam \\ 
Freq Range: & 24--87~MHz ($>$4:1 sky-noise dominated); 10--88~MHz usable \\ 
Instantaneous bandwidth: & $\le$16 MHz $\times$ 4 beams $\times$ 2 tunings \\ 
Minimum channel width: & $\sim$0 (No channelization before recording) \\ 
Beam FWHM: & [8,2]$^{\circ}$ at [20,80]~MHz for zenith-pointing \\ 
Beam SEFD: & $\sim$3~kJy (approximately frequency-independent) zenith-pointing \\ 
Beam Sensitivity: & $\sim5$~Jy (5$\sigma$, 1~s, 16~MHz) for zenith-pointing \\ 
All-Dipoles Modes: & TBN: 75~kHz bandwidth continuously from every dipole \\ 
& TBW: Full band (78~MHz) every dipole for 61~ms, once every $\sim$5 min.\\
\hline
\\
\end{tabular}
\end{center}
\end{table}

For the convenience of the reader, Table~2 provides a glossary of
defining some of the more arcane acronyms associated with LWA1.

\begin{table}[t!]
\centering
\caption{Glossary of LWA1 Terms}
\vspace{0.1cm}
\small
\begin{tabular}{ll}
Term & Description \\
\hline 
\hline
\vspace{-0.2cm}\\
CARC & The UNM Center for Advanced Research Computing\\
CFP & Call for Proposals\\
CGP & Crab Giant Pulses\\
DRSU & Data Recorder Storage Unit (a 10~TB eSATA drive array used by LWA1 data recorders)\\
IOC & Initial Operational Capability\\
LDA & LWA1 Data Archive (see \S4)\\
LEDA & Large aperture Experiment to detect the Dark Ages, a new backend for LWA1   \\
LSL & LWA Software Library; a collection of software for reading and analyzing LWA data\\
LWA & Long Wavelength Array; a future aperture synthesis radio telescope consisting of 53 \\
    & stations similar to LWA1\\
LWAC & LWA Collaboration\\
LWA1 & The first LWA station, now operating as a single-station radio observatory\\
LWANA & The LWA North Arm Stub Station located near the end of the VLA's North Arm (\S~3)\\
MCS & Monitor and Control System, the software and computers that control LWA1\\
PASI & Prototype All-Sky Imager (An existing backend for LWA1).\\
Stands & a pair of orthogonally-aligned active dipoles sharing a mast. \\
Station & An antenna array and associated electronics.  Analogous to a single dish, except \\
     & able to point beams in multiple directions simultaneously\\
TBN & Transient Buffer Narrowband (an LWA1 ``all sky'' observing mode)\\
TBW & Transient Buffer Wideband  (an LWA1 ``all sky'' observing mode)\\
\hline
\\
\end{tabular}
\end{table}

\begin{figure}[t!]
\begin{center}
\includegraphics[width=6.0in,angle=0]{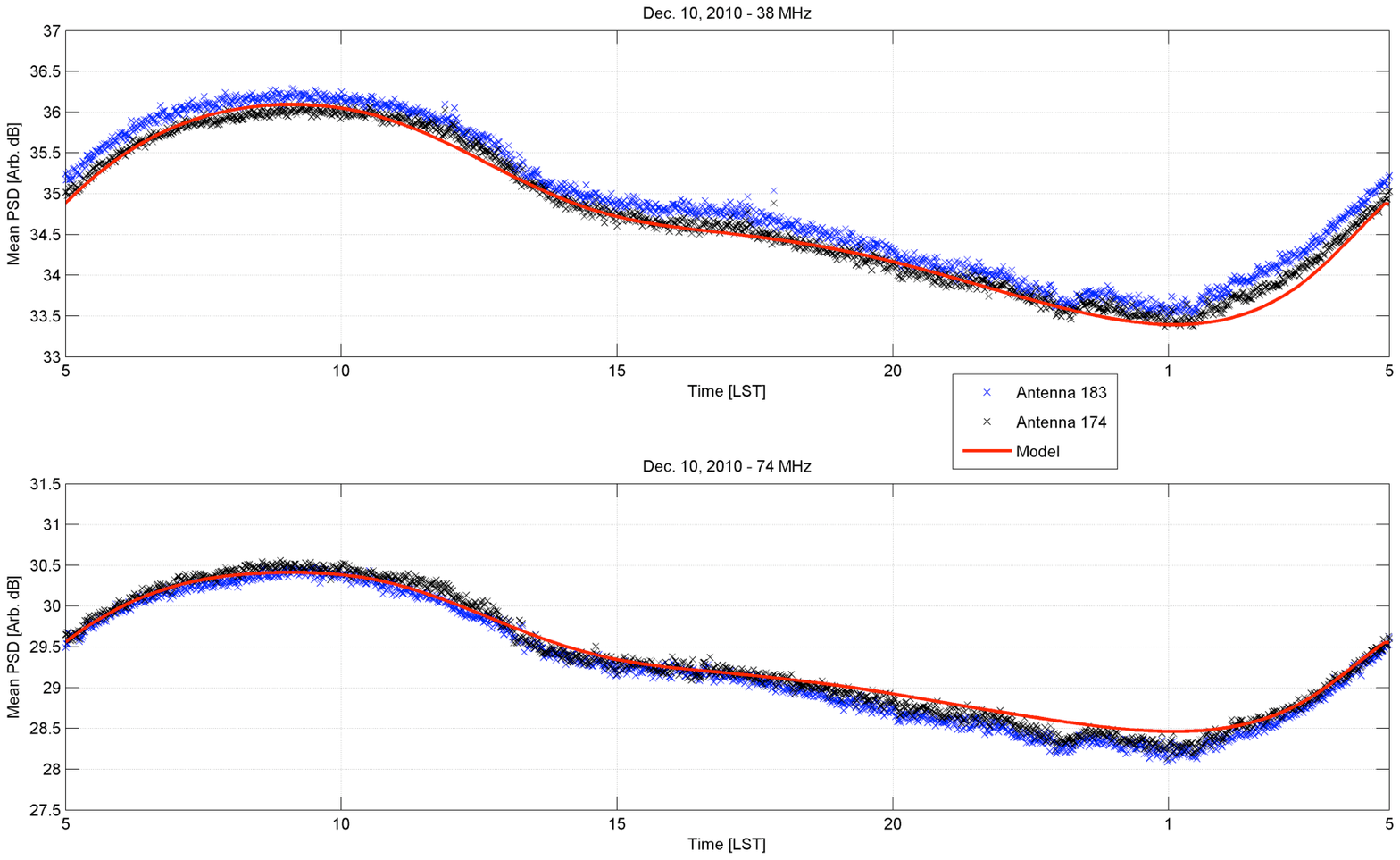}
\end{center}
\vspace{-0.5cm}
  \caption{
Dipole total power measurements (1 MHz bandwidth, 61 ms integration per point).  Variation
is due to the changing sky brightness temperature distribution as seen by the dipole.  The solid
red line is our prediction obtained by convolving the sky model of de Oliveira-Costa \etal\ (2008) with a model of
the dipole pattern obtained from electromagnetic simulation.  No RFI mitigation or editing
has been applied.
}
\label{fig:Dipole_Drift}
\end{figure}

At the time of this writing, science observations have begun
while commissioning continues.  We
anticipate reaching IOC (``initial operational capability'') --
essentially the beginning of routine operation as an observatory -- in
Spring 2012.  We now summarize some early results obtained during
commissioning. Fig.~\ref{fig:Dipole_Drift} shows single dipole total
power for period of 24 hours at 38 and 74 MHz for
two antennas.   The dipole power measurements compare well with a model of the 
expected power derived from a sky model convolved with the calculated
antenna pattern.  The agreement confirms that we are
strongly sky noise-dominated ($>$4:1 from 24-87 MHz), that we have a good understanding of our dipole
responses, and that RFI is manageable.

\begin{figure}[ht!]
\begin{center}
\includegraphics[width=5.2in,angle=0]{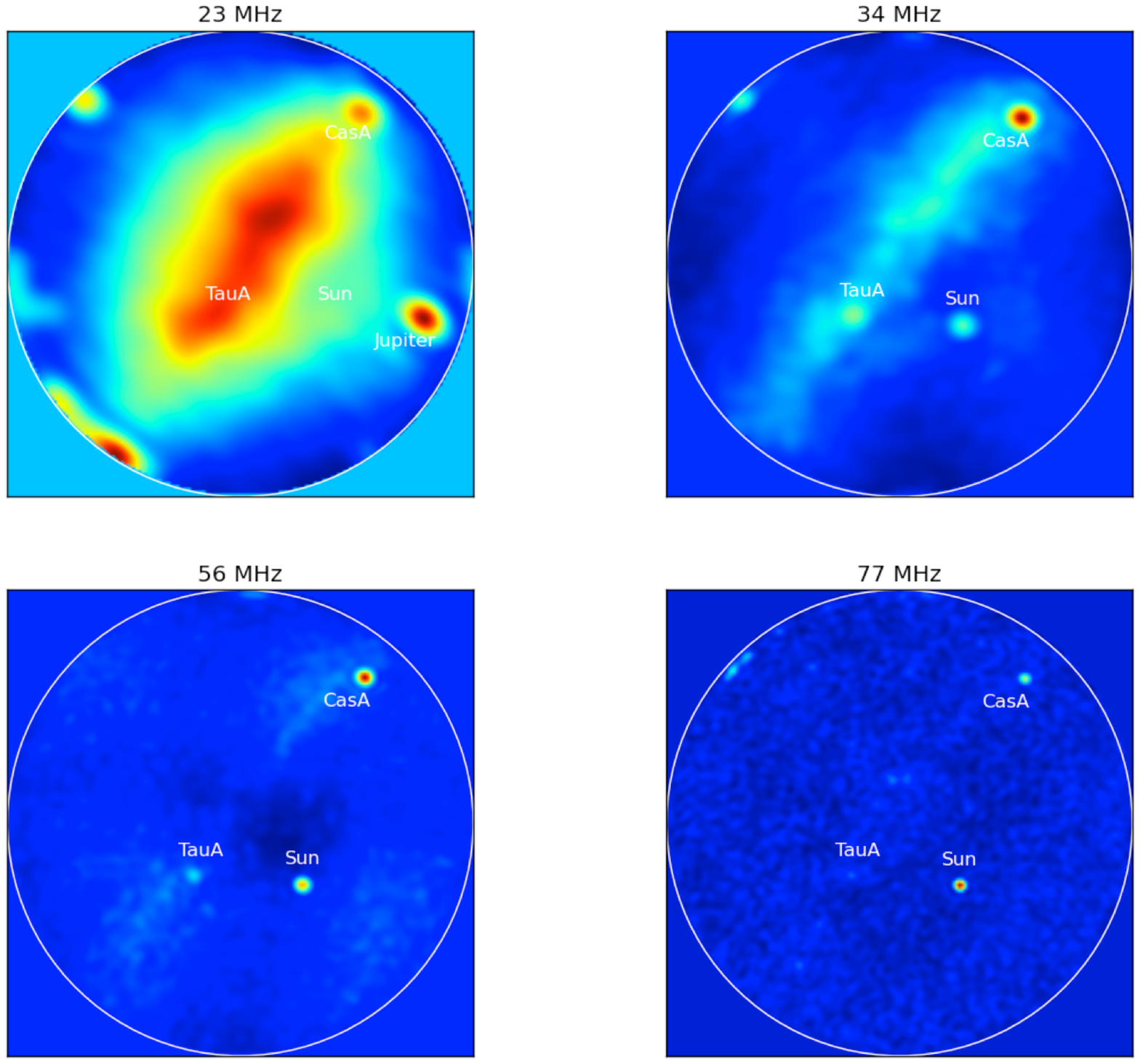}
\end{center}
\vspace{-0.5cm}
  \caption{
Nearly-simultaneous all-sky images taken at 4 widely separated frequencies
using LWA1's TBN mode.  Absolute calibration is the same in all four images;
the apparent decrease in sky brightness with increasing frequency is real, and
the bright region near zenith is the Galactic plane.  Clearly visible at 23 MHz
is Jupiter, and the horizon ``hot spots'' in the 23 MHz image  are ionospherically-refracted RFI.
Note that Cas A and the Sun are visible in all images.  
Data was obtained for 10 seconds each, 50 kHz bandwidth, using 
210 stands.
}
\label{fig:all_sky}
\end{figure}

We have begun imaging the sky with LWA1.
In Fig.~\ref{fig:all_sky} we show three views of the sky taken with 
the Transient Buffer Narrowband (TBN) mode on May 16, 2011 using 210 stands (21945 baselines).  In these 
Stokes I images one can see the
Galactic plane, Cas A, and Tau A, and at the lowest frequency 
Jupiter is quite prominent. LWA1 routinely images the sky
in near real-time using the Transient Buffer Narrowband (TBN) 
cabability of the station and a modest cluster located at LWA1.
These images are shown live on ``LWA-TV'' which is available from
the LWA web pages\footnote{\it http://www.phys.unm.edu/~lwa/lwatv.html}.  
Time-lapse movies of the
day's images are also made available at the end of each day.

\begin{figure}[htp!]
\begin{center}
\includegraphics[width=6.0in,angle=0]{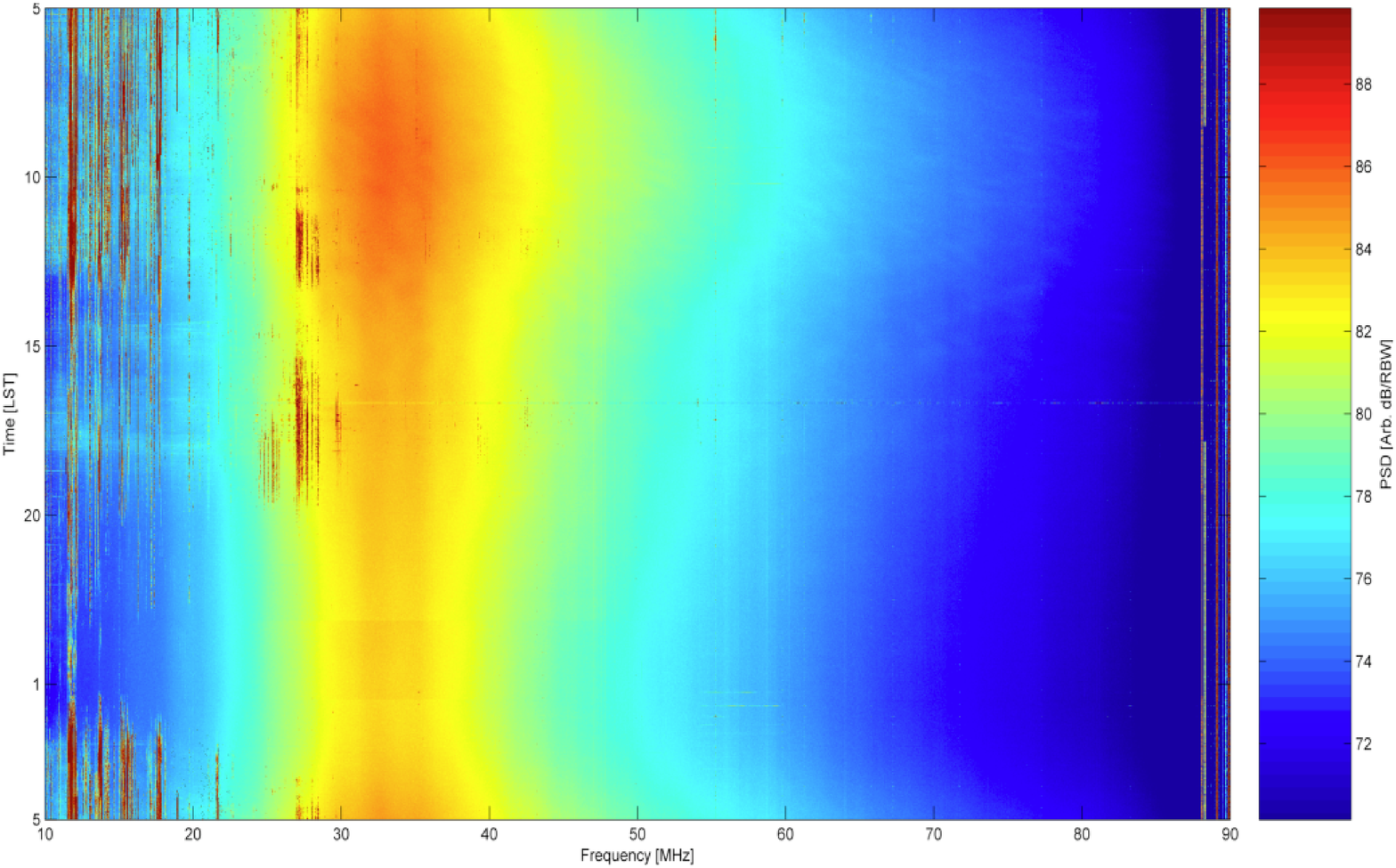}
\end{center}
\vspace{-0.5cm}
  \caption{
Spectrum using the TBW capture mode for 20 dipoles
phased at zenith for 24 hours.  The time and frequency variation of the 
background are real; the contribution of the active antenna appears
as a steep roll-off below 30 MHz. Note that 30-88 MHz is always
useable, and even frequencies as low as 13 MHz are usable for 
a few hours each day.
}
\label{fig:TBW}
\end{figure}


In Fig.~\ref{fig:TBW} we show a spectrogram obtained from TBW data
taken over 24 hours for a 20-dipole zenith-pointing beam.  The
integration time of the individual captures is 61 ms, and one capture
was obtained every minute.  The frequency resolution is $\sim$10 kHz.
The diurnal variation noted in Fig.~\ref{fig:Dipole_Drift} is apparent
here as well.  Strong RFI from the FM bands shows up as vertical lines
at 88 MHz and above.  Below 30 MHz there are a variety of strong
communications signals.  While there is abundant RFI visible in the
spectrum, it is very narrowband, obscures only a tiny fraction of our
band, and does not interfere with our
ability to be sky-noise limited.
More details about the RFI envirornment can be found in Obenberger \&
Dowell (2011).



\begin{figure}[ht!]
\begin{center}
\includegraphics[width=5.6in,angle=0]{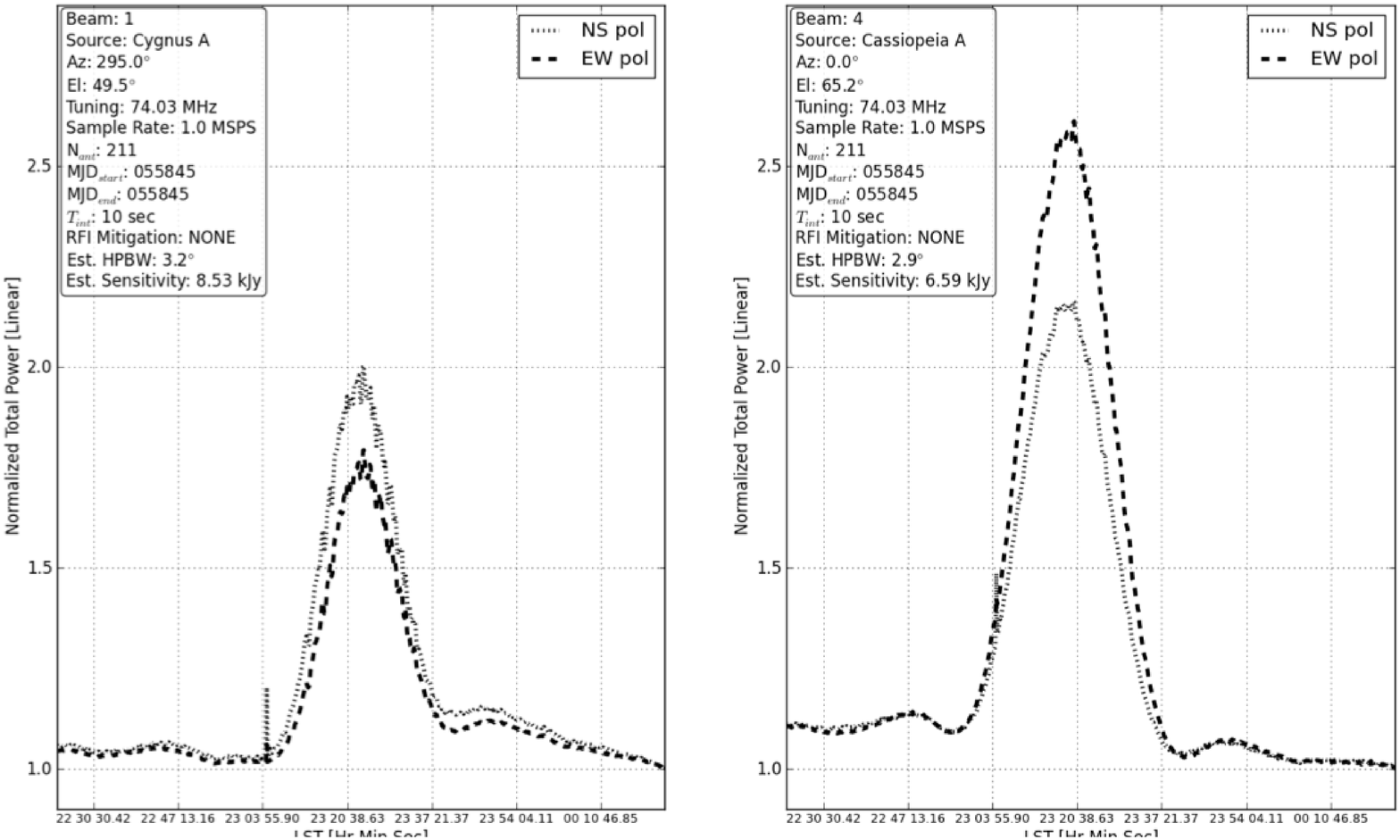}
\end{center}
\vspace{-0.5cm}
  \caption{
Simultaneous observations of Cyg~A and Cas~A using two 
of the four LWA1 beams at 74 MHz with a 1 MHz bandwidth.
}
\label{fig:multibeam}
\end{figure}

LWA1 supports the formation of up to 4 simultaneous beams.
In Fig.~\ref{fig:multibeam} we illustrate the formation of two
simultaneous beams, one placed on Cassiopeia~A and one placed
on Cygnus~A.  No tracking was performed so the two sources
drift through the beam.  
One can see the response of the beam
to Galactic emission as well as to Cygnus~A and Cassiopeia~A.  The two sources
have roughly equal flux density at 74 MHz (17 kJy; Helmboldt \& Kassim 2009), so the observed 
difference is primarily a result of the different dipole patterns 
at their respective elevations.

\begin{figure}[t]
\begin{center}
\includegraphics[width=4.2in,angle=0]{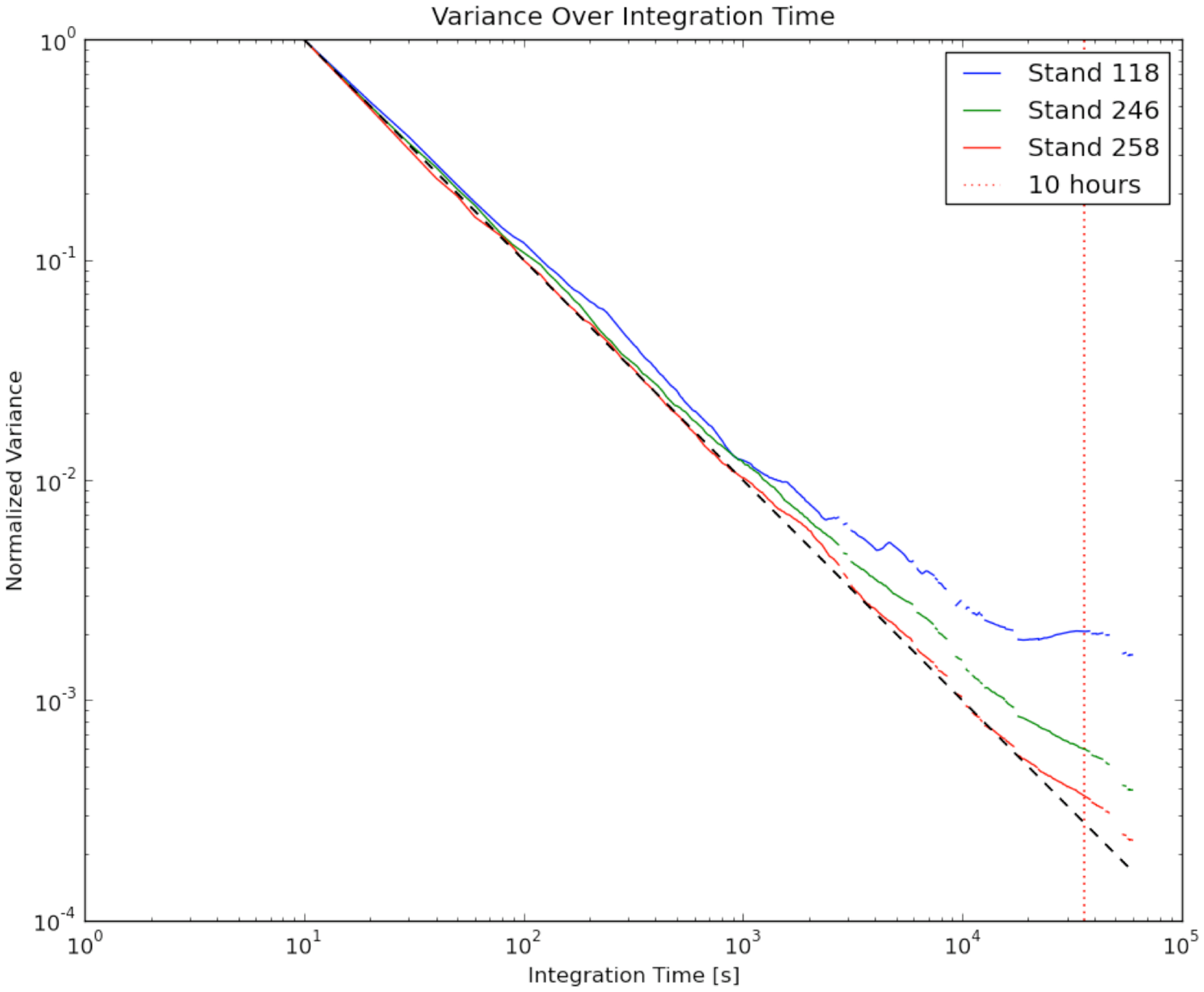}
\end{center}
\vspace{-0.5cm}
  \caption{
Variance over time using a $\sim$50 kHz bandwidth and 2048 channels at 38 MHz for three dipoles.  
The black dashed line shows the expectation of 
variance falling as the square root of integration time.  The vertical bar marks
10 hours.  Crude RFI excision was performed by discarding $\sim$20\% of the channels (in practice $<$1\% is generally sufficient).
}
\label{fig:deep}
\end{figure}

Figure~\ref{fig:deep} demonstrates that with very simple RFI mitigation,
LWA1's sensitivity is limited by noise alone -- as opposed to RFI or
instrumental stability -- for integrations up to many hours.  Note that we
have made no deliberate attempt in Fig.~\ref{fig:deep} to correct for 
instrumental stability (e.g., calibration against a noise standard), and 
that doing so can be expected to result in even better performance.

In Fig.~\ref{fig:B1919+21} we show the pulse profile from B1919+21, a strong, long period, low 
dispersion measure pulsar.  This was one of the first observations with the
beamforming mode of LWA1, and not only confirms the ability of LWA1 to do leading-edge
pulsar science, but also demonstrates that digital beamforming is working properly.  

\begin{figure}[t!]
\begin{center}
\includegraphics[width=4.7in,angle=-90]{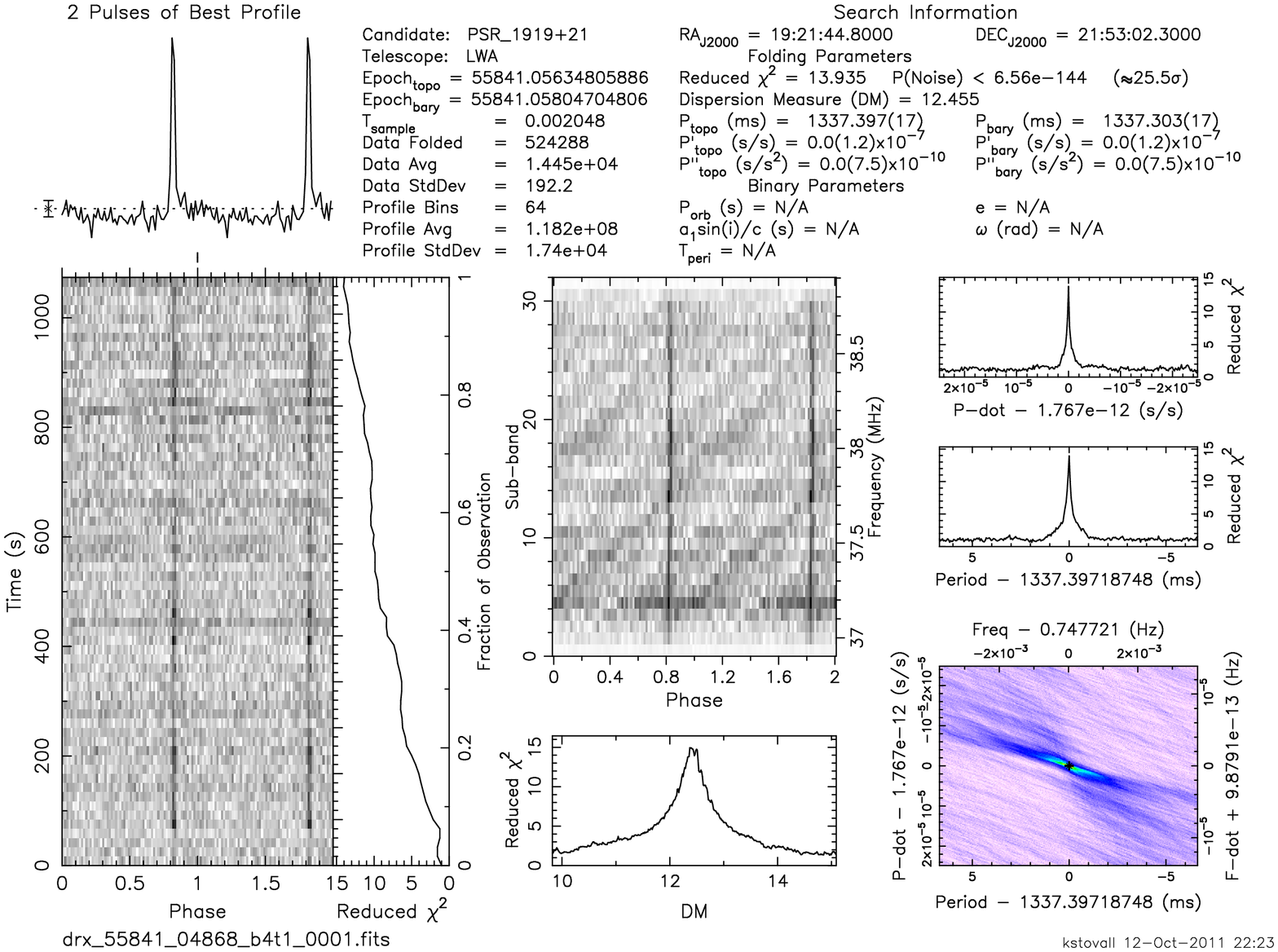}
\end{center}
\vspace{-0.5cm}
\caption{\label{fig:B1919+21} 
First LWA1 detection of the pulsar B1919+21 in a 1 MHz beam 
taken during commissioning observations.  The pulsar observations were
processed using PRESTO (Ransom 2001).  The panels are: (bottom left) 
plot of pulsar phase vs observation length averaged into 0.13 second bins;
(top left) integrated pulse profile dedispersed at optimal signal strength;
(top right) $\chi^2$ as a function of trial $\dot{P}$ at optimal $P$ and $DM$;
(middle right) $\chi^2$ as a function of trial $P$ at optimal DM and $\dot{P}$;
(bottom right) Reduced $\chi^2$ surface of trial $P$ and $\dot{P}$ at optimal $DM$;
(bottom middle) $\chi^2$ as a function of trial $DM$ at optimal $P$ and $\dot{P}$;
(top middle) Plot of pulse phase vs observing frequency. Data was folded at optimal $P$ and $\dot{P}$ for the entire
dataset for a particular sub-band (frequency range). In this case there are 
32 frequency sub-bands. These sub-bands are then plotted with dedispersion 
accounted for in order to show that the pulsar was seen throughout the 
observed frequency range.
}
\label{fig:B1919+21}
\end{figure}


\section{Scientific Program}

With the ability to point in several directions at once, wide
fractional bandwidths, and large collecting area, the first LWA
station will make important scientific contributions in several
fields.  We have been refining the science case for LWA1 over 5
years -- see Kassim \etal\ (2006, 2010), Clarke \etal\ (2009), Rickard
\etal\ (2010), and Lazio \etal\ (2010) for details.  Projects granted
observing time in the first (primarily internal) call for proposals
cover a range of topics including pulsars, transients, cosmology,
Galactic science, the Sun, and the Earth's ionosphere (see Table~3).
Below we elaborate upon several of these topics and the headway that
LWA1 will soon make in each.

\subsection{Pulsars} 

Pulsars are fascinating objects with spin periods and magnetic fields
strengths ranging over 4 and 5 orders of magnitude respectively.
Though it is well accepted that pulsars are rotating neutron stars,
the pulsar emission mechanism and the geometry of the emitting region
are still poorly understood (Eilek \etal\ 2005).  LWA1 will be an
excellent telescope for the study of pulsars including single pulse
studies, and studies of the interstellar medium (ISM).  In fact it is
in LWA1 frequency band range where strong evolution in pulsar
radio emission can be observed, e.g, a turn over in the flux density
spectrum, significant pulse broadening, and complex changes in the
pulse profile morphologies (Malofeev \etal\ 1994). And, although
pulsars were discovered at low frequency (B1919+21 at 81 MHz; 
Hewish \etal\ 1968), there is a remarkable
lack of observational data in the LWA1 frequency range. 
Fig.~\ref{fig:B1919+21} shows an LWA1 observation of B1919+21; and the 
LWA1 will detect dozens more (see Jacoby \etal\ 2007) in less than 1000
seconds.

LWA1 is able to perform spectral studies of
pulsars over a wide frequency range and with high spectral resolution.
This will allow investigators to look for drifting subpulses.  
Strong notches have been seen to appear in the profiles of 
pulsars at low frequencies (Hankins 1973), but little progress
has been made in understanding their origin.
Some pulsars may reach 100\% linear polarization at low frequencies
(B1929+10; Manchester \etal\ 1973).
In addition to being intrinsically of interest (providing clues
about the pulsar magnetospheric structure), such strongly 
polarized beacons can assist in probing coronal mass ejections
and determining the orientation of their magnetic fields.  That orientation
can strongly influence their effects if they impact the Earth (Kassim \etal\ 2010).

LWA1's large collecting area will be particularly useful for ``single pulse''
science, including studies of Crab giant pulses (CGPs; Bhat \etal\ 2007) and 
anomalously intense pulses (AIPs; Stappers \etal\ 2011).
The Crab Pulsar intermittently produces single pulses having intensity
greater than those of the normal periodic emission by orders of
magnitude. Despite extensive observations and study, the mechanism
behind CGPs remains mysterious. 
Observations of the Crab pulsar across the electromagnetic
spectrum can distinguish between various models for GP emission such as
enhanced pair cascades, radio coherence, and changes in beaming direction.
To date the study of the CGP emission at low radio
frequencies is only very sparsely explored.
Reported modern observations of CGPs in this frequency
regime are limited to 
just a few in recent years including UTR-2 at 23 MHz (Popov \etal\ 2006),
MWA at 200 MHz  (Bhat \etal\ 2007), and LOFAR LBA (Stappers \etal\ 2011).
LWA1 will be able to provide hundreds of hours per year of sensitive
observations of CGPs which could significantly increase our knowledge of the 
time- and frequency-domain characteristics of these enigmatic events.

We should be able to measure scattering for practically every good
CGP detection (S/N $\sim$20 or better), and it is known that both the
dispersion and scattering of the Crab emission can vary dramatically
on short or long time scales. By observing over an extremely
broad bandwidth, we may be able to better quantify the scatter
broadening and thereby assess the level and importance of
anisotropy. Furthermore, the broad bandwidth of the observations will
be helpful in shedding light on the issue of the frequency scaling of
the scattering (believed to be $\sim$3.6 compared to the canonical value
of $\sim$4.4 for the general ISM), which is thought to be related to the
nature of turbulence in the nebula.

Anomalous high intensity single pulses from known pulsars have been
reported previously using the UTR-2 (Ulyanov \etal\ 2006) and LOFAR (Stappers \etal\ 2011).  
These anomalously intense pulses (AIPs) have many features
similar to the giant pulse phenomenon, including emission in a narrow
longitude interval and power-law distribution of the pulse energy.  One
distinctive feature of these AIPs, however, is that they are generated by
subpulses or some more short lived structures within subpulses.  The
emission is seen to be quite narrow band, typically 1 MHz in bandwidth.
The nature of such pulses is not yet understood.  LWA1 with its
excellent sensitivity and large available bandwidth provides an
opportunity to study these pulses.

\subsection{Hot Jupiters} 

The magnetized planets in our solar system are well known to produce extremely
bright coherent radio emission at low frequencies (e.g., Carr \etal\ 1983).
In the past decade, with the abundant detection of planets around other
stellar systems, much theoretical work has been carried out on the detection
of similar radio emission from extrasolar planets (e.g., Lazio \etal\ 2009).
Such a detection is the only currently viable method for measuring the
magnetic fields and rotation rates of extrasolar planets, and it may provide
information about the other properties, such as interior composition.  For a
terrestrial-mass planet, a magnetic field may be important for determining its
habitability by protecting the planet from energetic particles.  The brightest
extrasolar planets are the Hot Jupiters (HJs), which we define as having
semimajor axes less than 0.5~AU and masses above 0.5~Jovian masses.

Using LWA1, we will conduct a volume-limited search out to 50~pc for
decametric emission from all known HJs in the northern hemisphere.  The
observational capabilities of LWA1 are uniquely suited to the properties
of these sources:
\begin{itemize}
\item {\bf Low-frequency coverage.}  Decametric emission from a HJ will
  exhibit a sharp cutoff at the electron cyclotron frequency of the planet's
  magnetic field, with typical cutoffs ranging from 1--100~MHz.  The LWA1's
  benign RFI environment allows observation down to 20~MHz routinely, and down
  to 10~MHz during favorable ionospheric conditions.
\item {\bf Sensitivity to polarized and ``bursty'' emission.}  HJ emission is
  expected to be highly polarized and released in bursts with timescales of
  $\sim$1~ms to $\sim$10~s; LWA1 observations provide full Stokes parameters
  and better than 1~ms temporal resolution.
\item {\bf Extensive coverage of many targets.}  The emission may be narrowly
  beamed, intersecting the Earth for a small fraction of the 10--100~hr
  rotational periods for some sources, while missing Earth entirely for
  others.  The multiple independent beams of LWA1 allow us to dedicate
  approximately 300~beam-hours to begin this survey.
\item {\bf mJy sensitivity.}  The large collecting area of LWA1 
allows us to achieve a typical (non-imaging) sensitivity of
  10~mJy for each observation in our survey (of duration 14 hours), lower than the $\sim$100~mJy flux
  densities predicted for some HJs (Griessmeier \etal\ 2007).
\end{itemize}
To date, searches below 70~MHz have had sensitivities of no better
than 100~mJy (Zarka 2007).  After only six months of observation, we will
either make a detection or be able to place tight constraints on models for
HJ emission.

Additionally, LWA1 will be used to conduct {\it blind} searches for HJ emission.  Based on
stellar density estimates of the solar neighborhood (e.g.,
Howard \etal\ 2010), we expect approximately 5,000 HJs within 100~pc.  At a
probable detection frequency of 20~MHz, an LWA1 beam is $\sim$50~deg$^2$ at zenith
(and larger still at lower altitudes), on average covering six nearby but
unknown HJs.  To detect these sources, we will add a search of bursty, highly
polarized emission to our  analysis of the beamformed data.
Candidate detections will be followed up with dedicated observational programs
to look for periodicity of such emission, a sign that we have detected a HJ or
similarly emitting brown dwarf.

\subsection{Transients} 

Astrophysical transient sources of radio emission signal the explosive
release of energy from compact sources (see Lazio \etal\ 2010, Cordes
\& McLaughlin 2003 for reviews).  Known types of radio transients
include cosmic ray airshowers, solar flares (\S2.5), Jovian flares and
flares from extrasolar hot jupiters (\S2.2), giant flares from
magnetars (Cameron \etal\ 2005), rotating radio transients (McLaughlin
\etal\ 2006), giant pulses from the Crab pulsar (\S2.1), and supernovae.  The
study of these sudden releases of energy allow us to recognize
these rare objects, and yield insights to the nature of the
sources including energetics, rotation rates, magnetic field
strengths, and particle acceleration mechanisms.  Furthermore, some
radio transients remain unidentified such as the galactic center radio
transient GCRT J1745$-$3009 (Hyman \etal\ 2005), and require further study.  A number of
sources have been predicted to produce strong radio transients, but
have not yet been observed, including prompt emission from 
gamma-ray bursts (GRBs; Benz \& Paesold 1998), neutron star 
mergers (Hansen \& Lyutikov 2001),
expiration of primordial black holes (Rees 1977; Blandford 1977),
topological phase transitions in primordial black holes (Kavic \etal\
2008), and superconducting cosmic strings (Vachaspati 2008).

A number of instruments have been built to study the meter-wavelength
transient radio sky including the Cambridge Low Frequency Synthesis
Telescope (Dessenne \etal\ 1996), the Fallbrook Low-Frequency
Immediate Response Telescope (FLIRT; Balsano 1999), the Eight-meter
Transient Array (ETA; Deshpande 2009), and the Long Wavelength
Demonstrator Array (LWDA; Lazio \etal\ 2010).  These surveys had
limited collecting area, field-of-view, or availability.  LWA1 will
far surpass the limits set by these instruments, in sensitivity by two
orders of magnitude, and with access to much of the sky all the time.

The Prototype All-Sky Monitor (PASI) is a software correlator and
imager for LWA1 that analyzes the TBN data stream, which provides
continuous samples from all dipoles with a 75~kHz passband placeable
anywhere within 10--88~MHz (see \S 3).  PASI images nearly the
whole sky ($\approx$$1.5\pi$~sr) every five seconds, continuously
and in near realtime, with full Stokes parameters and typical sensitivities
of $\sim$5~Jy at frequencies above 40~MHz and $\sim$20~Jy at 20~MHz.
Candidate detections can be followed up within seconds by beamformed
observations for improved sensitivity and localization.  These
capabilities provide an unprecedented opportunity to search the
synoptic low-frequency sky.  PASI 
saves visibility data for $\sim$20 days, allowing it to ``look back
in time'' in response to transient alerts. The images generated
by PASI will be archived indefinitely.

LWA1 will also search for single dispersed pulses or cosmic ``events''
using beamforming mode.  This mode trades all-sky capability for
greater bandwidth and time resolution.  Multiple beams will be used to
mitigate against interference.  In 50-70 MHz, any given coordinate
will remain in the beam for about 14 minutes, which is sufficient to
view the entire duration of pulses with DM up to $\sim$1100 pc
cm$^{-3}$.


Ultimately, we plan for LWA1 to generate and accept alerts for
multi-wavelength follow-up; its observations will be even more
valuable in conjunction with the results of other facilities surveying
large areas of the sky (e.g., Fermi, Pan-STARRS, LSST, LIGO).  
Furthermore, LWA1 is
complementary to LOFAR in that the two instruments have little
simultaneous overlap on the sky, but complimentary in that it has
much better access to the Galactic center and inner plane where
one might expect some transient populations to be concentrated
(Kassim \etal\ 2003).

\subsection{Cosmic Dawn and LEDA} 

Understanding the origin, formation, and evolution of the first
galaxies is one of the major questions of astrophysics. Identified by
the Astro2010 decadal review (New Worlds, New Horizons in Astronomy and 
Astrophysics) as an area ripe for exciting new
discoveries, the period of ``Cosmic Dawn'' encompasses the formation
of the first galaxies and black holes. These luminous sources produce
UV and X-ray radiation that radically alters the properties of the
diffuse neutral hydrogen gas that fills the majority of space at these
times. It is crucial that we develop observational probes of this era
since this would give valuable insights into the energetics of the
early universe and the processes of initial galaxy formation.  One
unique avenue for observationally constraining the radiative
properties of the first luminous objects is via low frequency radio
observations of the redshifted 21 cm line of neutral hydrogen (Madau
\etal\ 1997).

\begin{figure}[t!]
\includegraphics[width=3.0in,angle=0]{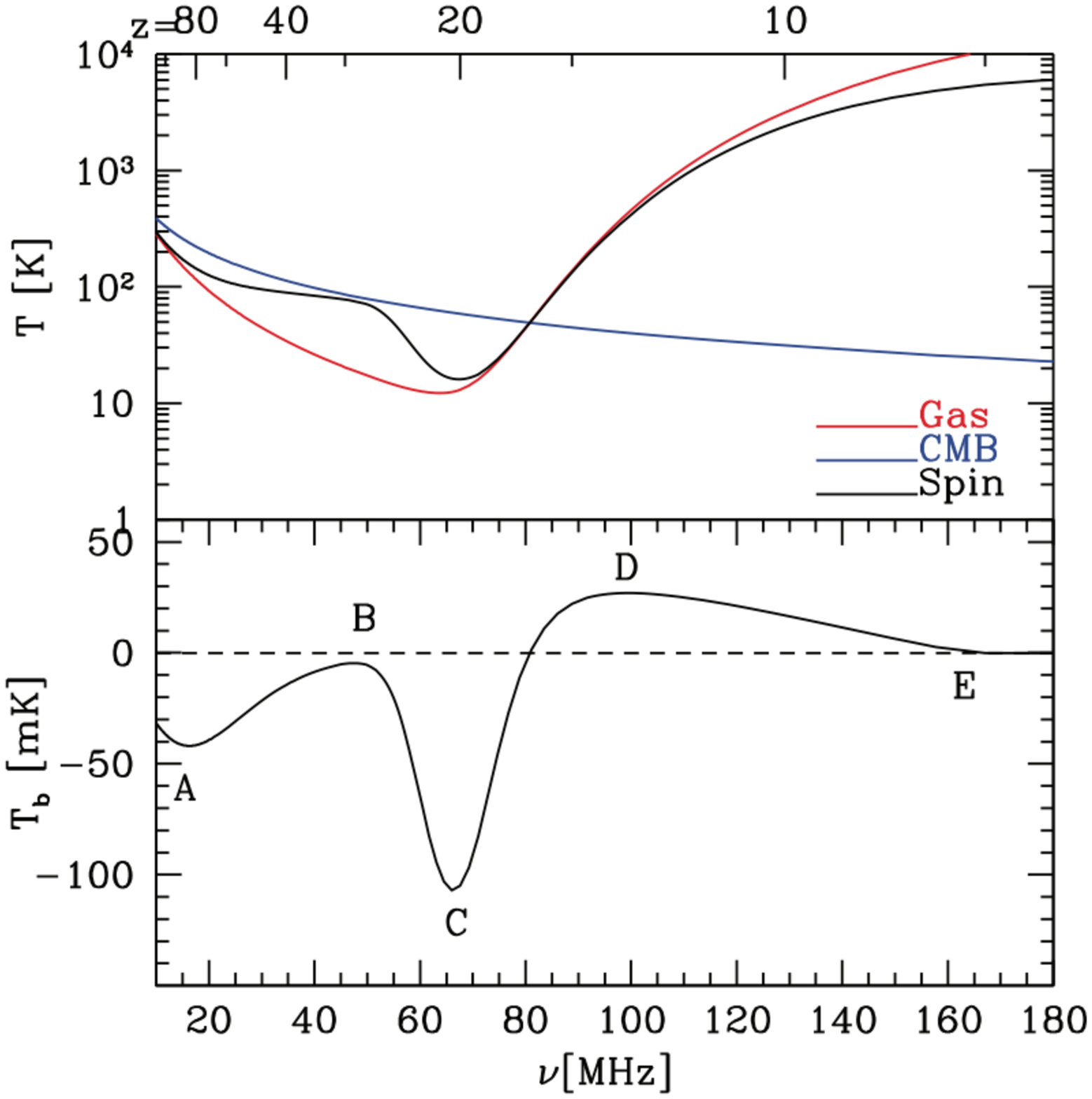}
\hspace{0.1cm}
\includegraphics[width=3.0in,angle=0]{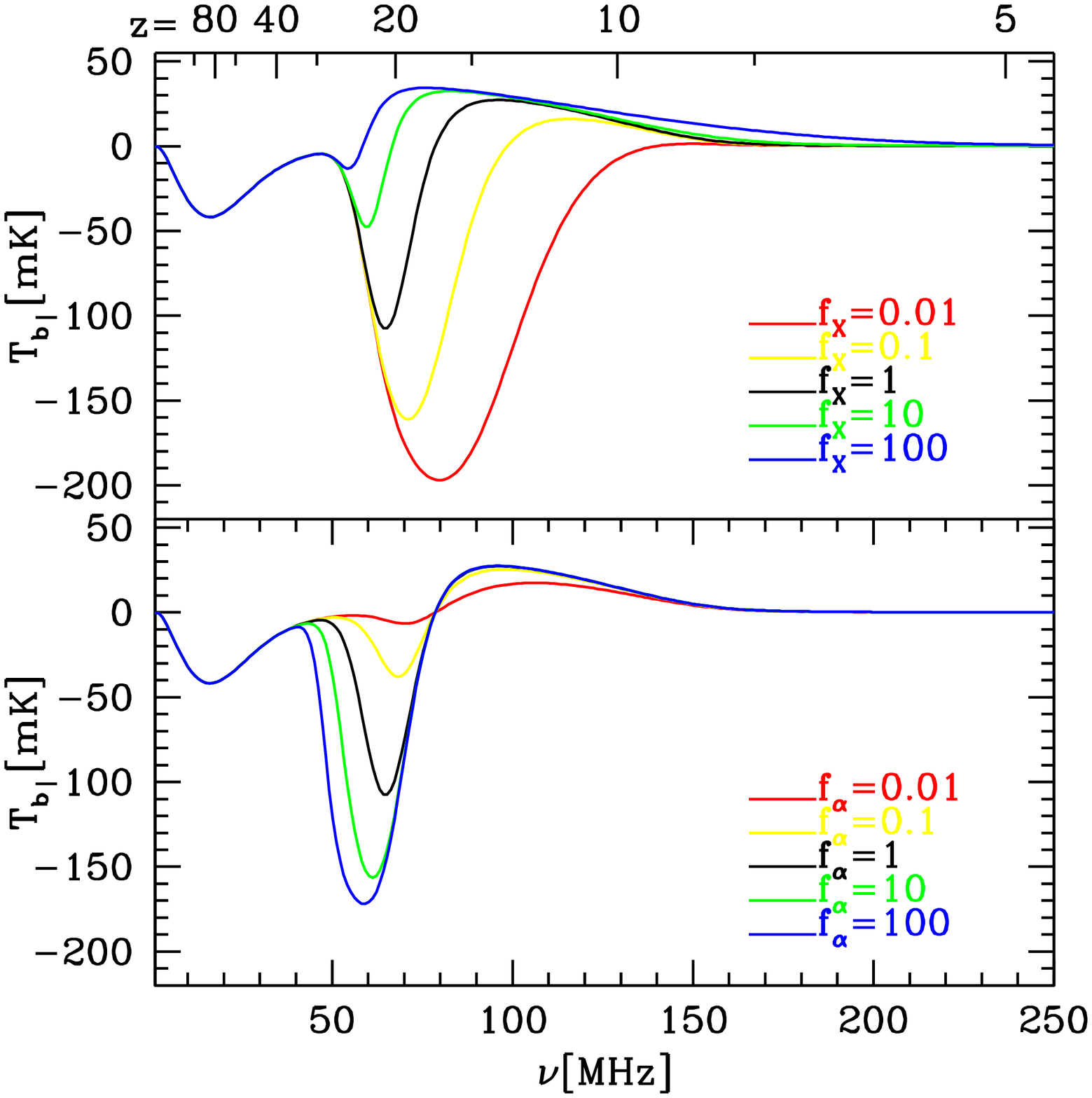}
\caption{\label{fig:darkages} Evolution of temperature scales 
relevant for the 21 cm signal as a function 
of frequency. Top left panel: Evolution of the 
gas (red curve), CMB (blue curve), and spin 
(black curve) temperature for the fiducial 
history of Pritchard \& Loeb (2010). Bottom left panel: Evolution of the 21 cm brightness temperature with inflection points (A,B,C,D,E) labeled. 
At (A) collisional coupling of the gas and spin temperature begins to become ineffective; at (B) Ly$\alpha$ photons begin to couple the gas and spin temperatures producing an absorption feature; at (C) X-rays from stars and/or galaxies heat the gas and eventually lead to an emission signature which peaks at (D); finally reionizaton removes all signal at (E).  
LWA1 will probe the regions from 20-84 MHz where the signal is in absorption. 
Right: Alternative scenarios for 21 cm heating from X-rays (top) and
Ly$\alpha$ (bottom) which
changes the strength of the absorption signal.  Here $f_x$ is the product
of the X-ray emissivity and the star formation efficiency, and $f_a$ is the 
product of the Ly$\alpha$ emissivity and the star formation efficiency.
}
\end{figure}

Coupling with the gas temperature results in the HI line being seen in
absorption at high redshifts, when the gas is cooler than the CMB due
to adiabatic expansion (during the so-called ``Dark Ages''); it
appears in emission at lower redshifts, after the gas becomes heated
by X-rays from the first stars and black holes. According to
theoretical models, the turnover between these regimes is around z
$\sim$ 17, at 80 MHz (Furlanetto \etal\ 2006; Pritchard \& Loeb 2008),
and the maximum absorption occurs between 50 and 100 MHz (Pritchard \&
Loeb 2010).  Fig.~\ref{fig:darkages} shows a theoretical model for the
temperature history of the early universe, along with the predicted HI
absorption signature.

J. Bowman and collaborators have proposed a novel approach to
measuring the global 21 cm signature using LWA1.  They will detect
or constrain the absorption peak that marks the end of the Dark Age
using LWA1. LWA1 is well-matched to the expected frequency range
of the absorption peak, and by simultaneously forming a science beam
that targets a relatively cold region of sky and a strong source with
a known spectrum, they will achieve a lower foreground and a better
spectral calibration than single-antenna experiments
allow. Furthermore, they will be able to repeat the experiment using
different patches of sky, which further reduces the problem of
foregrounds and allows for follow-ups of any marginal detections. This
approach requires no additional hardware beyond what has already been
developed for LWA1 and can be accomplished in $\sim$500 beam
hours.

Greenhill, Werthimer, Taylor, and Ellingson have proposed a more
ambitious method to measure or constrain the HI signal from the
transition between the Dark Ages and widespread reionization. The
Large Aperture Experiment to Detect the Dark Ages (LEDA) 
\footnote{\tt http://www.cfa.harvard.edu/LEDA/}
comprises (i)
a large-N correlator deployed to the LWA1 (512 inputs, 60 MHz), (ii)
hardware to equip four outrigger dipoles for (redundant) calibrated
total-power measurement, and (iii) a data calibration pipeline with
which to reconstruct the full-Stokes sky brightness over a $\sim
140^\circ$ field of view, measure and correct for direction-dependent
gains of individual stands, and measure and correct for refractive
offsets created by the ionosphere (see Greenhill \& Bernardi 2012).
In LEDA, the array will serve to enable calibrations that cannot be
achieved with a single antenna; the outriggers will deliver the data
streams from which the 21cm science will be derived.  The wide-band,
wide-field sky model and antenna calibrations derived from array data
will enable foreground subtraction (from the total power data).  Use
of multiple outriggers and a calibration array that can be subdivided
to yield independent measurements will enable quantification of
systematics.
Cross-correlation for a large-N array scales as O(N$^2$) and poses
significant computing challenges. Hundreds of apertures and frequency
channels, and tens of MHz bandwidth require 10$^{13}$ to 10$^{14}$
operation per second.  The LEDA FX correlator will combine Field
Programmable Gate Arrays (FPGAs) serving the O(N) F stage and Graphics
Processing Units (GPUs) serving the O(N$^2$) X stage (Clark, LaPlante,
Greenhill 2012). Application of GPUs, which may be coded in a C-like
environment, reduced development time and cost while delivering nearly
80\% utilization of the GPU floating point resources (single
precision) and ready reconfigurability.

We note that detecting the 21 cm signal from the Dark Ages has
many challenges to overcome including foreground subtraction, 
understanding sidelobe systematics, mitigating RFI and mutual coupling,
and coping with ionospheric distortions.  It could be that space-borne
measurements are needed to bypass RFI and ionospheric effects.  
The Dark Ages Radio Explorer (DARE) mission (Burns \etal\ 2012) has
been proposed to deploy a radio telescope into an orbit around
the moon to measure the Dark Ages signal.

\subsection{Solar Bursts} 

\begin{figure}[t]
\begin{center}
\includegraphics[width=5.0in,angle=0]{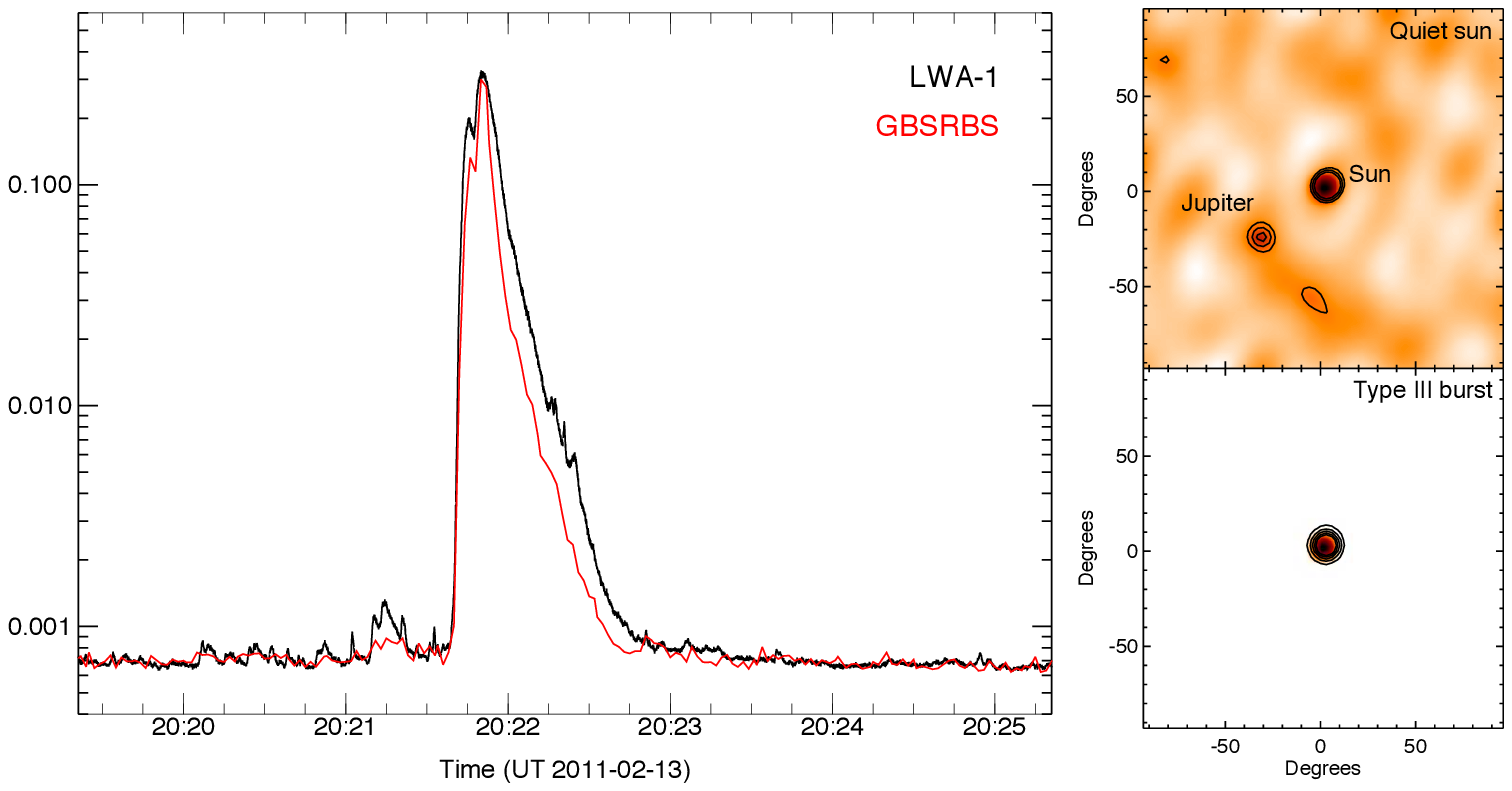}
\end{center}
\vspace{-0.5cm}
\caption{\label{fig:type3} 
(Left) The 30 MHz light curve of a solar Type III burst at 0.04 s 
time resolution (bandwidth 50 kHz) from LWA1 (black line), compared with (red 
line) the 2 s time resolution, 2 MHz bandwidth light curve from the 
Green Bank Solar Radio Burst Spectrometer (GBSRBS; White \etal\ 2005). The noise level in LWA1 data is 
much lower than in the GBSRBS data (from a single dipole antenna), and 
the low-level LWA1 variability is real. (right) Comparison of LWA1 maps of the 
sky centered on the Sun before the burst (upper, 2011 Feb 13 at 20:19:20 UT) 
and near the peak of the burst (lower, 20:21:50 UT). Both maps contain a few 
seconds of data and use 4 kHz of bandwidth. The maps were made using data from 
just 8 dipoles relatively close to each other in LWA1. The resolution is 
10 degrees. The peak in the burst map is about 5000 times brighter than the 
peak in the pre-burst map.}
\end{figure}

LWA1 will use single-beam measurements of the Sun, taking
advantage of the sensitivity and temporal and spectral resolution
available with LWA1, to look specifically at the fine structure in Type
II and Type III bursts, and possibly to track moving Type IV
bursts. The combination of sensitivity and time resolution will permit
us to use the wave field statistics to test proposed models for the
emission process.  An example of this is shown in Fig.~\ref{fig:type3}
taken during the solar event on 2011 Feb 14 using the TBN mode with 10
dipoles and the prototype LWA1 digital processor.

Type III bursts are generated by beams of nonthermal keV-energy electrons
propagating out from the Sun on open field lines, producing radio
emission at the local plasma frequency by plasma emission, i.e.,
conversion of electrostatic Langmuir waves into electromagnetic
waves. Since the electrons are travelling rapidly outward into 
a plasma of decreasing density, the
corresponding radio bursts drift rapidly downwards in frequency and
appear as nearly vertical features in frequency-versus-time plots
(``dynamic spectra''). The currently favored 
``stochastic growth'' theory for plasma emission predicts a  
log-normal distribution for wave-field strengths in these bursts
(e.g., Cairns \& Robinson 1999) that can be tested with LWA1 data. 
By contrast, Type II bursts are associated
with shocks propagating through the solar corona. They are usually
associated with coronal mass ejections, which can generate shocks as
they move through the solar atmosphere at high speed. However, there
are aspects of Type II bursts that appear to be correlated with the
structure of the flaring region itself.  Type II bursts show much
substructure in frequency-time plots that may be associated with
inhomogeneity in the shock, and with the mechanism for 
generation of Langmuir waves. The
exact nature of the source of plasma emission in Type II bursts is
still not entirely understood: it could be due to widespread electron
beams, as in Type III bursts, but this is more likely to be true of
the structures known as ``herringbone'', short-lived fast-drift
structures seen to emerge from the backbone of Type II emission. LWA1
data sampled at high temporal and spectral resolution may give clues
to acceleration processes in the shock creating the burst.
Moving Type IV bursts
are associated with large solar eruptions and can move several degrees at 
thousands of km/s: in extreme cases LWA1's high signal-to-noise should 
permit it to measure such motions even with the large beam
and reinvigorate a field of study that has been largely dormant for 20 years
due to the lack of imaging at LWA1's frequencies where these intriguing 
(and potentially space-weather relevant) phenomena occur. 


%

\subsection{Radio Recombination Lines} 

Radio recombination lines (RRLs) arise in ionized and partially
ionized gas, and therefore offer valuable probes of the physical
condition of the diffuse ionized interstellar medium. In particular,
the size of the electron orbit increases for high quantum numbers,
making the particles extremely sensitive to temperature and density of
the medium. At LWA1 frequencies the excitation temperature approaches
the kinetic temperature and RRLs are expected to be seen in
absorption, confirmed by carbon RRL detections (Fig.~9; Stepkin et
al.\ 2007; Kantharia \& Anantharamaiah 2001; Erickson \etal\ 
1995). Hydrogen RRLs at these frequencies are as yet undetected, perhaps
due to a dielectronic electron capture mechanism increasing the chance
of absorption in atoms with multiple electrons such as carbon (Walmsley
\& Watson 1982).

Our knowledge of carbon (and other species) RRLs is currently
relatively limited, and LWA1 opens up a new venue for observations
of these lines. In particular, the versatile bandwidth and spectral
resolution will be essential for successful observations of RRLs (Peters
\etal\ 2011). 
For example, at frequencies $\sim$26~MHz the spectral resolution required
to separate between the carbon RRLs is $\sim$125~kHz (see Fig.~9). At
the same time, a large bandwidth is desired since then many lines can
be observed at the same time, and the resulting spectra can be folded
to search for a possible detection. Using three beams at LWA1 we
can simultaneously cover about 85 carbon transitions with a 0.5~kHz
spectral resolution (6 km/s). This is straightforward for LWA1 because
the native format for data recording is voltage samples, as opposed 
to spectrometer output; channelization can be performed in software
at any desired frequency resolution.
The 100-m aperture size of LWA1 also offers improved angular resolution over 
single-dish studies (Erickson et al. 1995).

The first approved RRL observing project (see Table~3) includes
proof-of-concept observations of the known lines in Cas A to work out
any necessary calibration and reduction procedures. Once these are
established, three targets in the Cygnus Arm have been selected as
suitable places for an initial RRL search. The three targets are DR21
(HII region), HB21 (supernova remnant) and DR4 (supernova remnant),
all of which are bright at low radio frequencies and are likely to
contain large amounts of ionized or partly ionized gas. Future
possible and obvious RRL studies with LWA1 include surveys of large
regions to map the conditions of the diffuse ISM, and searches for
other RRL species.

\bigskip
\begin{figure}[ht]
\hspace*{2cm}\resizebox{12cm}{!}{\includegraphics{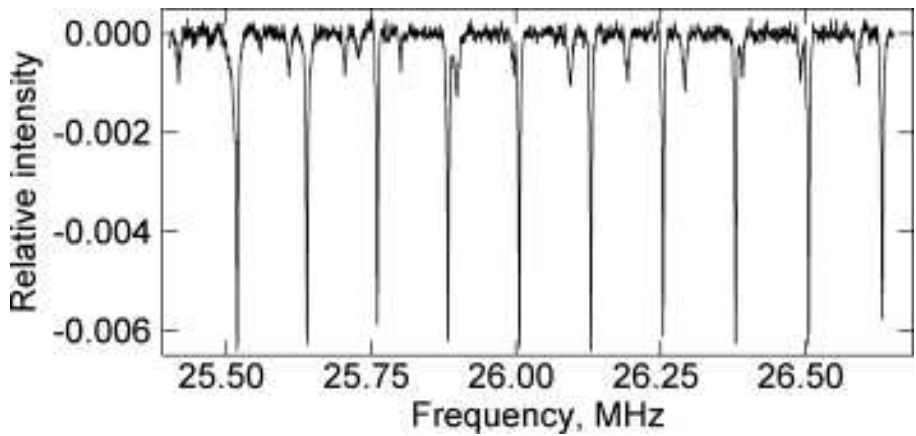}}
\label{fig:spec}
\caption{The spectrum of carbon at 26~MHz taken in the direction of Cas A
  (Stepkin \etal\ 2007)}
\end{figure}

%
 

\begin{table}[ht!]
\begin{center}
\caption{LWA1 approved observing projects}
\vspace{0.1cm}
{\scriptsize
\begin{tabular}{llllllll}
 \hline\noalign{\smallskip}
 \hline\noalign{\smallskip}
Project Title & Beam  & TB     & PI & PI & Co-I   \\
                      & hours & hours  &   & Affil. &  \\
\hline\noalign{\smallskip}
 LWA Cosmic Ray Air Shower Trigger &    & 144 & D.\ Besson & KU & UNM  \\
 Tracking the Dynamic Spectrum of  & 56 & 20   & T.\ Clarke & NRL & MTSU,UNM, \\
  Jupiter                           & & & & & NRAO,UTas  \\
 Ionospheric Scintillation         & 344 & & P.\ Crane & NRL & UCSD  \\
 Passive Meteor Scatter using      & 40  & 20 & S.\ Close & Stanford & PSU,Stanford \\
  the Long Wavelength Array         & & & & & \\
 A GCN-Triggered Search for        & 320 & & S.\ Ellingson & VT & \\
  GRB Prompt Emission               & & & & &  \\
 Crab Giant Pulses                 & 640 & & S.\ Ellingson & VT & JPL,NRL,Swinburne \\
 Continuing Measurements of the    & 6   & & J.\ Hartman & NRL & \\
  Cas~A/Cyg~A Flux Ratio            & & & & & \\
 Searching for Hot Jupiters with   & 300 & & J.\ Hartman & NRL & UCB \\
  LWA1                         & & & & &  \\
 Carbon Radio Recombinations Lines & 378 & & Y.\ Pihlstr\"om & UNM & NRL,UTas \\
  in the Cygnus Arm                 & & & & & \\
 Multi-frequency Large Scale Sky   & 672 & & E.\ Polisensky & NRL & UTas  \\
  Survey                            & & & & & \\
 Low Frequency Studies of Radio    & 50  & & P.\ Ray & NRL & VT,NMT \\
  Pulsars                           & & & & &  \\
 Ionospheric Absorption Measurements&432 & 216 & L.\ Rickard & UNM & JPL,NRL \\
  using LWA1 as an Imaging Riometer & & & & & \\
 Single Dispersed Pulses           & 320 & & J.\ Simonetti & VT & UMd,CWRU,CNJ \\
 Observing the Transient Universe  & & Cont. & G.\ Taylor & UNM & VT,NRL, \\\
   with the First LWA Station        & & & & & JPL, LANL \\
Solar Radio Bursts at High        & 320 & 160 & S.\ White & AFRL & UNM,NRL,GMU,  \\
Temporal and Spectal Resolution   & & & & & UMd,UTas & & \\
LEDA                              & 300 & & L.\ Greenhill & CfA & \\
Observing Cosmic Dawn with LWA1    & 520 & & J.\ Bowman    & ASU & UNM \\
  \noalign{\smallskip}\hline \\
\end{tabular}
}
\end{center}
Notes: These projects have already been allocated observing time in 2011/2012.
\end{table}

\section{Observatory Description}

\begin{figure}[t]
\vspace{0.5cm}
\includegraphics[width=6.0in,angle=0]{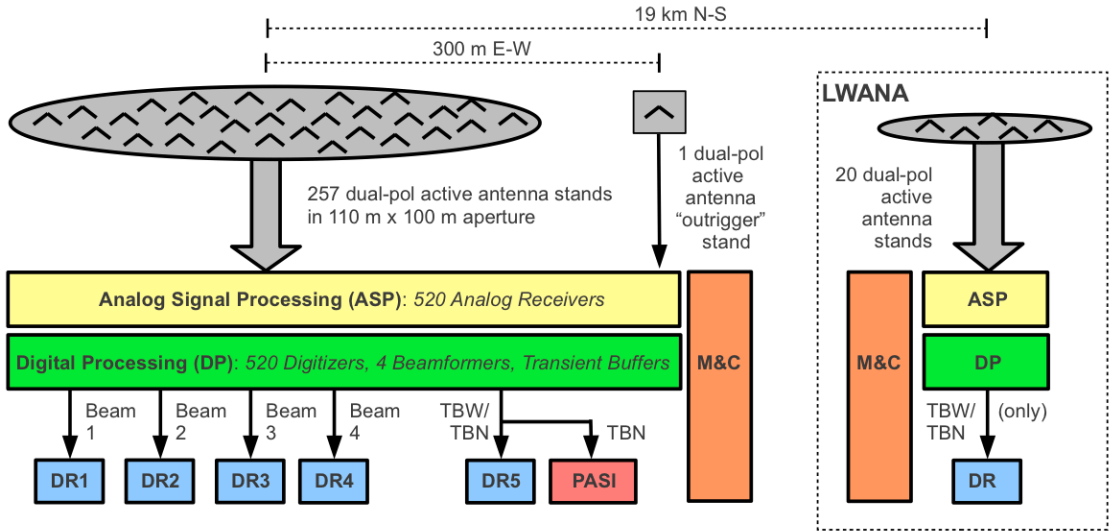}
\vspace{-0.5cm}
\caption{\label{fSysArch} LWA1 system architecture, simplified for clarity;  
details available in Craig (2009).}
\end{figure}

LWA1 lies on NRAO property, just a few hundred meters southwest of
the center of the VLA (see Figure~\ref{fig:lwa_design}).  The LWA1
system architecture is shown in Figure~\ref{fSysArch}.  The array
consists of 258 active antenna stands.  All but
two of the antennas are located within an ellipse of 100~m
(East--West) $\times$ 110~m (North--South), with the axial ratio
intended to improve beam shape for pointings toward the Galactic
Center, which transits at $Z \approx 63^{\circ}$ as seen from the
site.  The other two stands are outliers (one 77 m SW of center, and one
300 m E of center) used primarily for
calibration and diagnostic purposes.  Additional outliers are planned
as part of the LEDA project.

\begin{figure}[t]
\includegraphics[width=3.0in,angle=0]{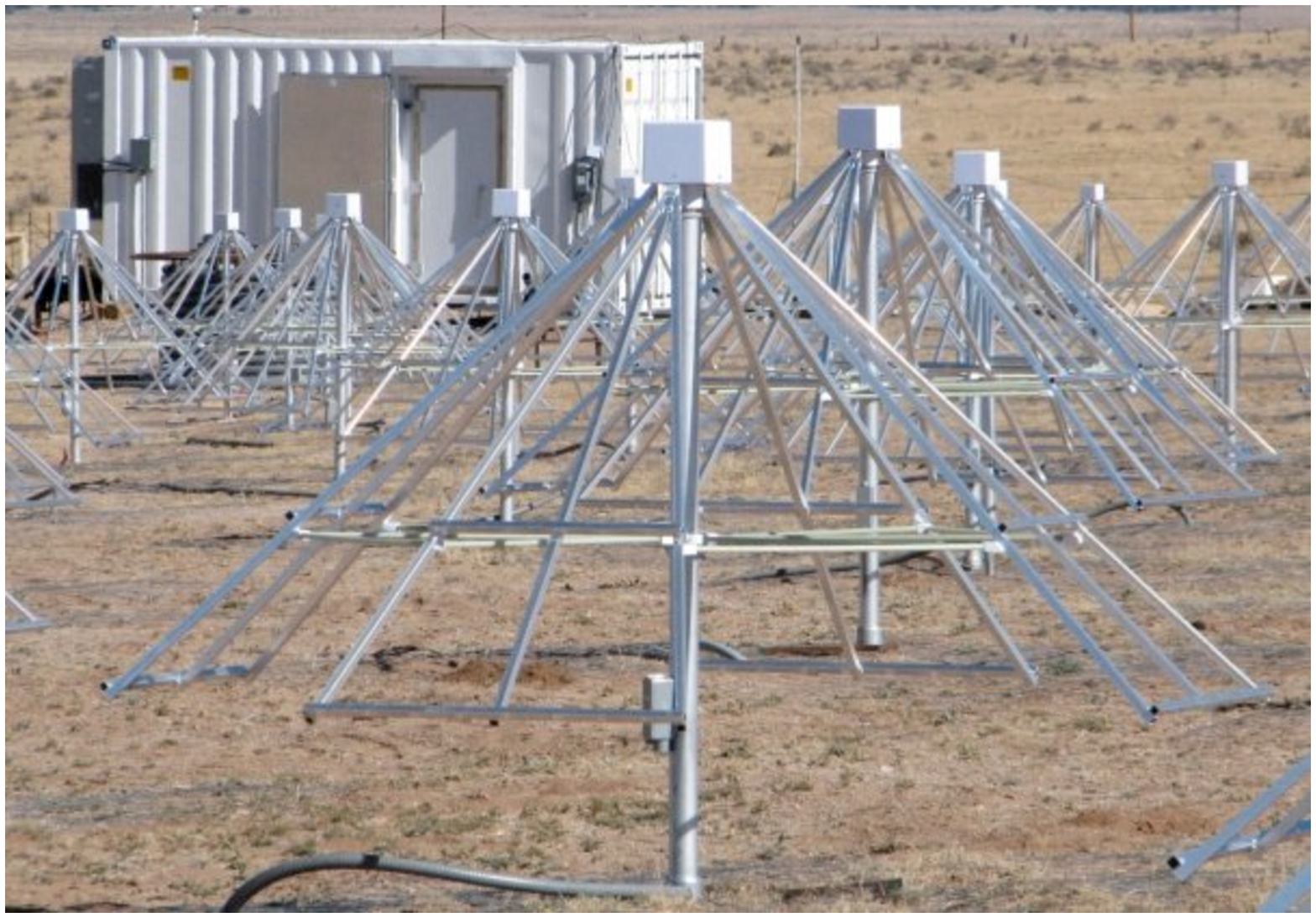}
\hspace{0.1cm}
\includegraphics[width=3.0in,angle=0]{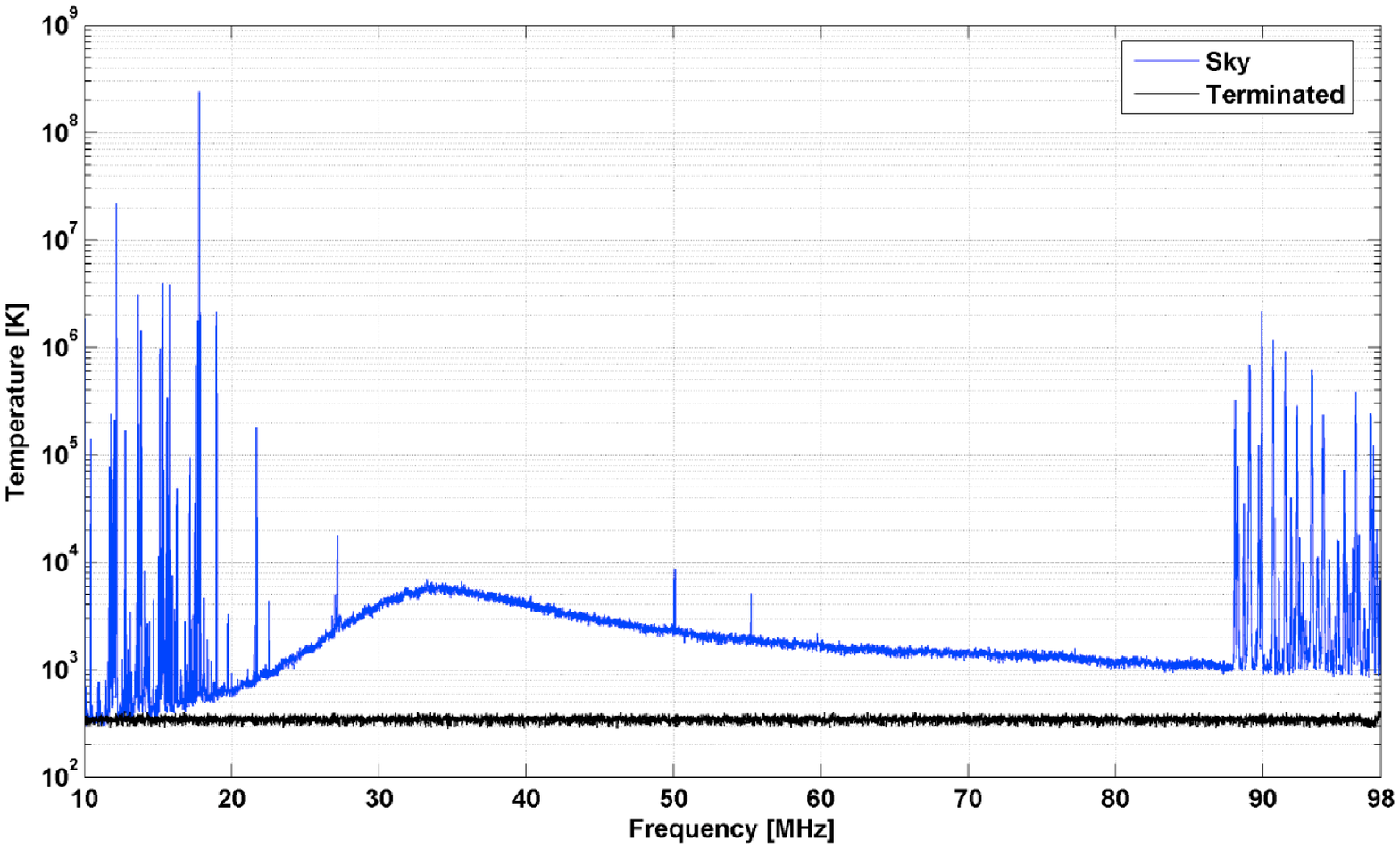}
\caption{\label{fStd} LWA1 active antenna. (a) An active antenna ``stand'', consisting of two orthogonally-aligned bowtie-type dipoles on a mast, over a 3~m $\times$ 3~m ground screen. (b) Output spectrum of one polarization of the active antenna and output with antenna terminals shorted, confirming that the output is sky noise-dominated over most of the 10--88~MHz tuning range. }
\end{figure}

\begin{figure}[ht]
\includegraphics[width=2.8in,angle=0]{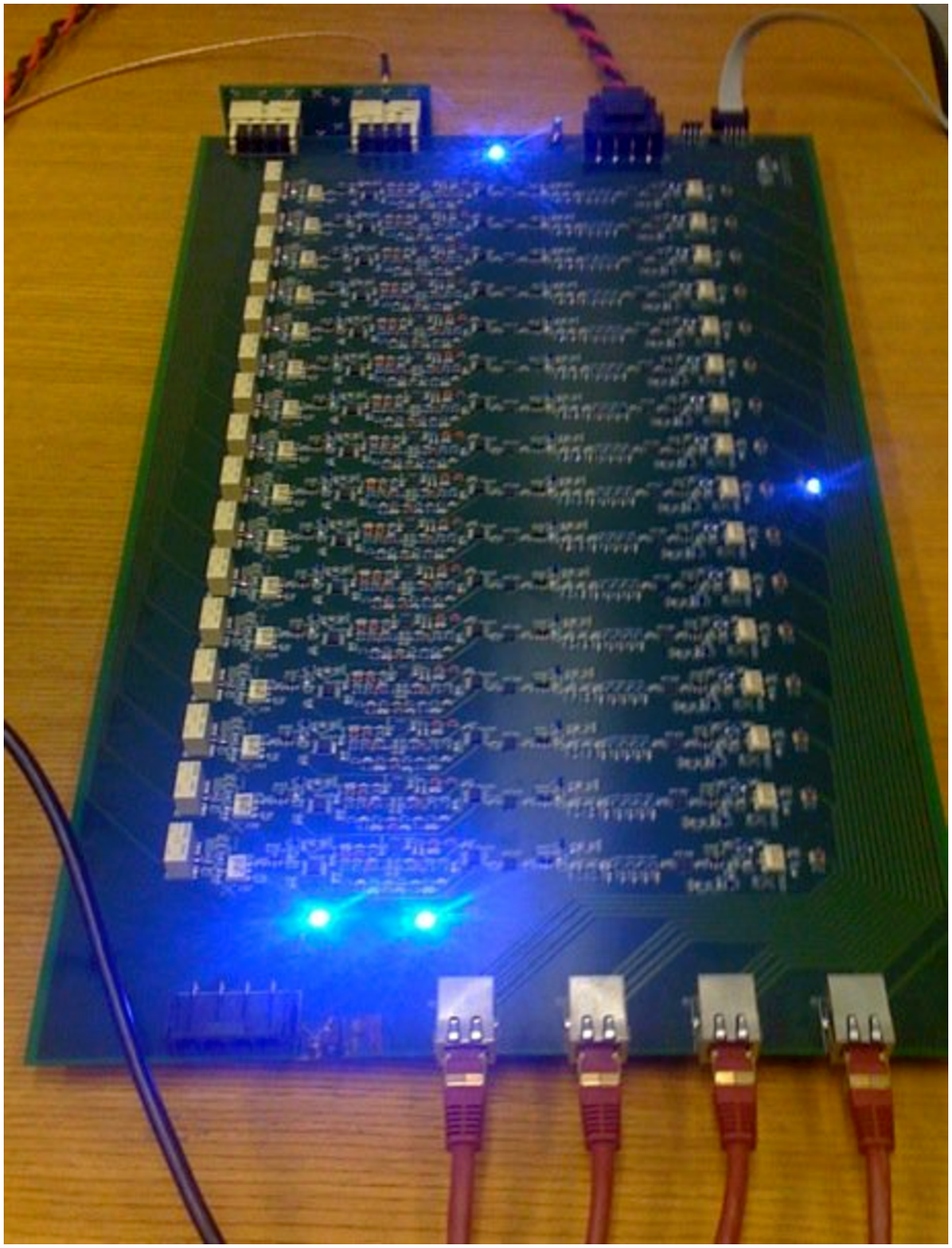}
\hspace{0.1cm}
\includegraphics[width=3.25in,angle=0]{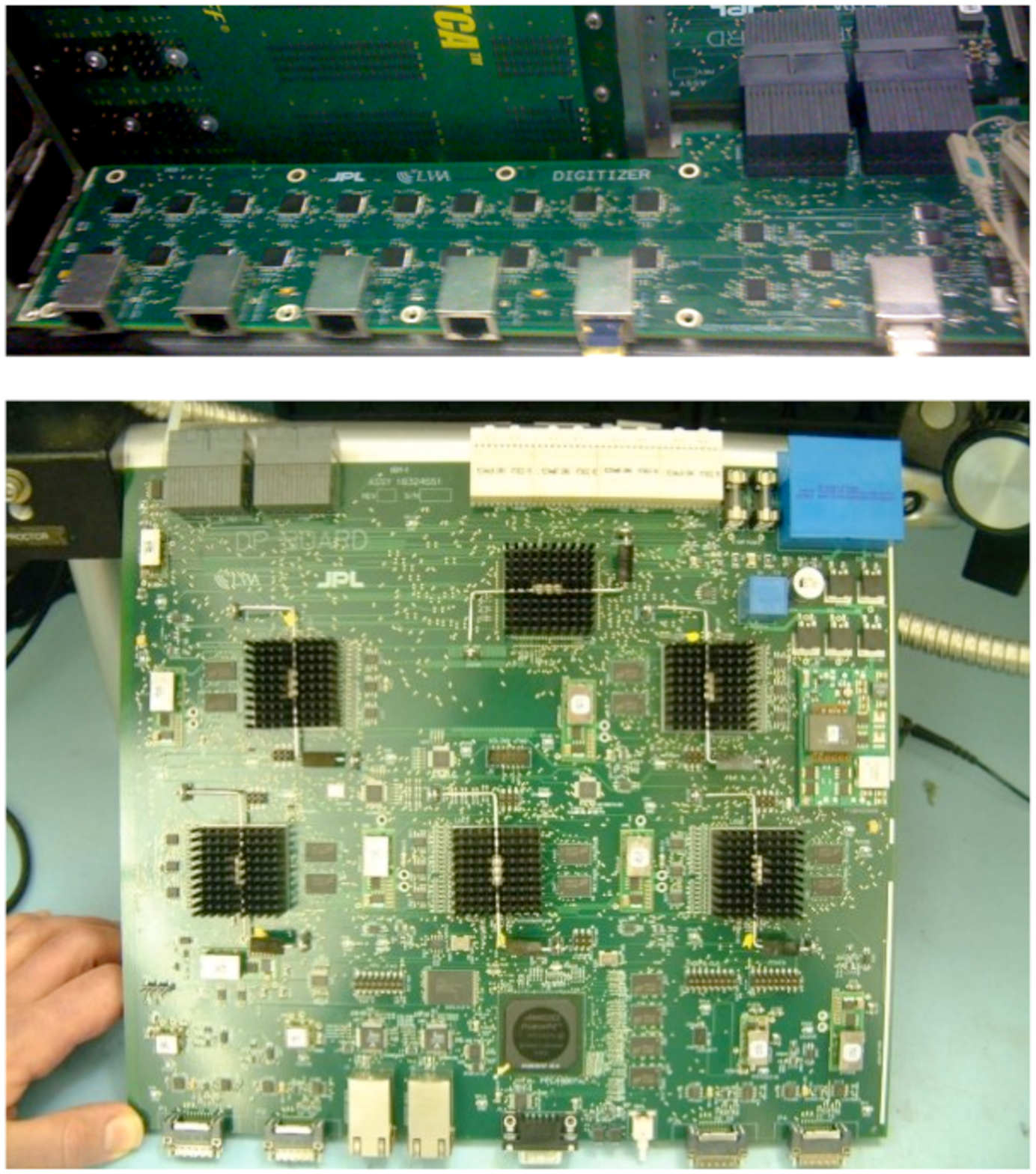}
\caption{\label{fARXDP} Left: One of LWA1's 33 16-channel analog receiver boards. Top Right: One of the digitizer boards; Bottom Right: One of LWA1's 26 20-channel ``DP1'' digital processing boards. Note the 5 FPGAs on the DP1 boardswhich are Xilinx Vertex V series. }
\end{figure}

An LWA1 antenna stand is shown in Figure~\ref{fStd}a.  Integral to
the stand is an ``active balun'' that provides about 35~dB of gain to
overcome subsequent losses, ensuring that the system temperature is
dominated by Galactic ({\it not} internal) noise, as demonstrated in
Figure~\ref{fStd}b.  Upon arrival in the shelter, the signal from
every antenna is processed by a direct-sampling receiver consisting of
an analog receiver, a 12-bit analog-to-digital converter (A/D) which
samples 196 million samples per second (MSPS), and subsequent digital
processing (``DP'') (See Figure~\ref{fARXDP}b).  This choice of sample
rate ensures that strong RFI from the 88--108~MHz FM broadcast band
aliases onto itself, with no possibility of obscuring spectrum in the
10--88~MHz tuning range.  Beams are formed using a fully-digital
time-domain delay-and-sum architecture, which allows the entire 78~MHz
passband associated with each antenna to be processed as a single data
stream (Soriano \etal\ 2011).  Delays are implemented in two stages: An
integer-sample ``coarse'' delay is applied using a first-in first-out
(FIFO) buffer operating on the A/D output samples, followed by a
28-tap finite impulse response (FIR) filter that implements an all-pass
``subsample'' delay.  The filter coefficients can be also specified by
the user, allowing the implementation of beams with custom shapes and
nulls.  The delayed signals are added to the signals from other
antennas processed similarly to obtain beams.  Four dual-polarization
beams are constructed in this fashion, each fully-independent and
capable of pointing anywhere in the visible sky.  Each beam is
subsequently converted to two independent ``tunings'' of up 
to $\approx$16~MHz bandwidth each, with each tuning having a center
frequency independently-selectable from the range 10--88~MHz.  Both
tunings of the beam emerge from DP as a single stream of UDP packets
on 10~Gb/s ethernet.  Thus there is one 10~Gb/s ethernet output cable per
beam (or ``pointing'').

Each beam output is connected to a data recorder, a PC that
records the UDP packets to a ``data recorder storage unit'' (DRSU).  A
DRSU is five 2~TB drives (10~TB total) in a 1U rack-mountable chassis,
configured as an eSATA drive array.  Each data recorder can host up to
2 DRSUs.  At the maximum beam bandwidth, each data recorder has a
capacity of $\approx$72~hrs of observation which is continuous except
for 1 gap of approximately 5~minutes, needed for the switchover
between DRSUs.  DRSUs may be either taken from the shelter for
analysis, or they may be offloaded onto other DRSUs or
commercially-available external USB hard drives.  It is also possible to transfer
data from data recorders directly off-site using the internet; however
the limited data rate of the internet connection to the LWA1 site
makes this impractical for observations longer than a few minutes.

Independently of beamforming, and as demonstrated in 
Figures~2, 3 and 4, LWA1 is able to coherently
capture and record the output of all its A/Ds, where each A/D
corresponds to one antenna.  This can be done in two distinct modes.
The ``transient buffer -- wideband'' (TBW) mode allows the raw 12-bit
output of the A/Ds to be collected continuously, but only for $61$~ms
at a time, and another 5 minutes is required to write out the captured
samples.  The ``transient buffer -- narrowband'' (TBN) mode, in
contrast, allows a single tuning of $75$~kHz bandwidth to be recorded
continuously and indefinitely, resulting in an output data rate of
$\sim$112~MB/s for the array.  These two modes share the
same 10~Gb/s ethernet output from DP, and thus are mutually exclusive.
However, the TBW/TBN output is distinct from the four beam outputs
and runs simultaneously with all four beams using a
dedicated (fifth) data recorder and DRSUs.  Although intended
originally to support commissioning and diagnostic functions, the TBW
and TBN modes have emerged as popular modes for science observations,
as indicated in \S~2.
 
LWA1 includes a dedicated backend known as the Prototype All-Sky
Imager (PASI).  PASI shares the TBN/TBW output and operates $\sim$80\%
of the observatory uptime. PASI is a cluster of 4 rack mounted servers
with Nahalem multicore processors interconnected by an Infiniband
switch.

Separate from LWA1, we have built the LWA North Arm
stub station (LWANA), a separate station located at the end of the
VLA's North Arm (approximately 19~km North of LWA1) primarily using spares
left over from the original procurements of components for LWA1.
LWANA consists of 20 stands (thus, $\sim8$\% of the collecting area of
LWA1) and is capable of TBW and TBN only (not beamforming).  However,
it is identical to LWA1 in all other respects, including data
recording.  

{\bf Radio frequency interference (RFI) environment:} As demonstrated
in Figures \ref{fig:Dipole_Drift}, \ref{fig:TBW}, \ref{fig:deep}, and \ref{fStd}b, the
RFI environment at LWA1 is surprisingly benign.  The observatory
management team has had extensive experience -- and consistent success
-- observing at these frequencies at this site using LWA1, a
precursor system (LWDA; Lazio \etal\ 2010), and the VLA 4-m system
(Kassim \etal\ 2007a).  This is not to say that observing in 
10--88~MHz is without challenges.  RFI impacts LWA1 on two levels:
First, by creating a potential threat to the linearity of the active
antennas and analog receivers; and second, by obstructing spectrum of
interest.  We consider the linearity issue first.  At LWA1, the most
troublesome interference is due to distant transmitters at frequencies
below about 30~MHz, whose signals arrive at the station by refraction
from the ionosphere, followed closely by FM broadcast signals in the
88--108~MHz range.  These linearity challenges are met using
high-intercept point active front end electronics in which we have
great confidence (the same design with only minor modifications has
been in nearly continuous use at the site since 2006).  Also the LWA1
analog receivers (Fig.~\ref{fARXDP}a) can be electronically reconfigured between three
modes: A full-bandwidth (10--88~MHz) uniform gain mode, a full-bandwidth dual-gain mode in which frequencies below
about 40~MHz can be attenuated using a ``shelf filter'' (useful for
observing above 40~MHz during that part of the day when long-range
signals in the 10--30~MHz band are strongest); and a 28--54~MHz mode,
which would serve as a last line of defense should we find in the
future that RFI above and/or below this range becomes persistently
linearity-limiting.  In addition, the total gain in each bandpass mode
can be adjusted over a 60~dB range in 2~dB steps, allowing fine
adjustments to optimize the sensitivity-linearity tradeoff.

Concerning the obscuration of science signals from in-band RFI: Our
extensive experience in this frequency range (from the 74~MHz VLA
system, from LWDA and ETA, and from LWA1 itself) has demonstrated
that RFI from both external and internal sources will inevitably be
present at all levels and throughout the spectrum.  We occasionally
encounter bouts of crippling interference, which we inevitably find to
be internally generated, from local powerlines, or due to activity at 
the nearby VLA
facilities.  In each case we have (with the excellent
cooperation from the VLA RFI management team) quickly identified and
eliminated the source.  For the remaining low-level RFI, we find that
the simple time \& frequency editing techniques currently employed are
effective in dealing with interference encountered most of the time.
In fact, we typically find that the primary difficulty posed by RFI is
not really that it prevents science, but rather that it increases the
amount of manual effort required to reduce data, and decreases
somewhat the amount of data that is ultimately useful.  

We plan a continuing program of technical development aimed
at characterizing, expanding and improving the capabilities of LWA1
and, in the process, training students in the art of instrument
development and developing the engineering expertise needed to tackle
the next generation of large meter-wavelength radio telescopes.  The 
LWANA (see \S~3)
is available as a testbed for technical development projects, so that
work can proceed with less concern about inadvertently degrading LWA1
with unproven modifications, or limiting its availability to
accommodate testing.  Technical developments planned include:
(1) On-the-fly data processing using the existing data recorders;
(2) Flexible scheduling, triggering, and rescheduling of
  observations;
(3) Enhanced and automated techniques for mitigation of RFI; 
(4) Beams with custom shapes and/or optimized sensitivity; and
(5) Improved sidelobe and polarization characterization.

\section{Observatory Operations}

\subsection{Allocation of Observing Time}

LWA1 observing time is allocated in units of ``beam-hours'' and
``TB-hours''.  One beam-hour corresponds to the dedicated use of one
of LWA1's four independently-steerable beams for one hour.  Similarly,
one TB-hour corresponds to the dedicated use of LWA1's TBW/TBN mode
for one hour.  Data will be shipped to users on hard drives, or made
available on-line through the LWA Data Archive (LDA), a 24 TB 
on-line archive hosted at UNM's Center for Advanced Research in
Computing.
  
The first LWA1 CFP was conducted in October 2009, and resulted in
the time allocations identified in Table~3.
CFPs will be issued twice annually.  
LWA1 also supports ``target of opportunity'' observing. The Director (or
in his absence, the Associate Director or Chief Scientist) is
authorized to grant up to 16 beam-hours (sufficient to cover the full
tuning range of LWA1 continuously for one pointing, for 4 hours)
and 8 TB-hours in response to a well-justified email request from
anyone.   

Finally, it should be noted that the LDA will
provide, in effect, a retroactive observing capability that we will
also make available to the community.

\subsection{The LWA Software Library}

The LWA project has developed a suite of software
tools for the scheduling and analysis of LWA data (the LWA Software
Library -- LSL, Dowell et al. 2012; and see\footnote{\tt http://fornax.phys.unm.edu/lwa/trac/wiki}), but we do not have the 
resources to provide turn-key software solutions for all
experiments that could be envisaged.  We have
established a clearinghouse for LWA Software and this will
be available to all users.  LSL initially benefited from
a software repository developed by NRL for the LWDA, and software
development at Virginia Tech.
LSL runs under Linux and also 
on Intel Macs.  Currently LSL includes readers for all 
LWA observing modes, correlators for the TBW and TBN 
dipole modes, and various scripts to generate plots of
antenna locations, spectra, spectrograms, and 
more.  LSL is publicly available from the LWA web
pages. 

\section{Conclusions}

LWA1 offers considerable sensitivity and sky coverage in the
relatively unexplored frequency range 10 -- 88 MHz.  LWA1 is
particularly interesting from an education perspective in that the
TBN/TBW modes provide one of the first ``Large-N'' radio telescopes,
challenging existing algorithms.  Furthermore, there are a number of
other low frequency instruments under commissioning and development
(LOFAR, MWA, PAPER) which face similar challenges in RFI mitigation,
calibration, and wide-field imaging.

The astronomical community is invited to apply
for time on this new facility at one of the upcoming proposal
deadlines.  

{\bf Acknowledgements:}

We acknowledge the efforts of the following students and postdocs who
helped to design or build LWA1: Sunil Danthuluri, Albino Gallardo,
Sudipta Ghorai, Aaron Gibson, Eduard Gonzalez, Mahmud Haun, Aaron
Kerkhoff, Ted Jaeger, Kyehun Lee, Justin Linford, Qian Liu, Adam
Martinez, Frank Schinzel, Abirami Srinivasan, D.W. Taylor III, Chenoa
Tremblay, Steve Tremblay, Sushrutha Vigraham, Chris Wolfe, Jayce
Dowell, Jake Hartman, Bryan Jacoby, Ted Jaeger, and Nagini Paravastu.
Basic research in radio astronomy at the Naval Research Laboratory is
supported by 6.1 base funding.  GBT acknowledges support from the
Lunar University Network for Astrophysics Research (LUNAR)
(http://lunar.colorado.edu), headquartered at the University of
Colorado Boulder, funded by the NLSI via Cooperative Agreement
NNA09DB30A.  Part of this research was carried out at the Jet
Propulsion Laboratory, California Institute of Technology, under a
contract with the National Aeronautics and Space Administration.  The
Centre for All-sky Astrophysics is an Australian Research Council
Centre of Excellence, funded by grant CE11E0090.  Construction of the
LWA has been supported by the Office of Naval Research under Contract
N00014-07-C-0147. Support for operations and continuing development of
the LWA1 is provided by the National Science Foundation under grant
AST-1139974 of the University Radio Observatory program.


\clearpage

\begin{thebibliography}{plainnat}



\bibitem{bals}
Balsano, R.J. 1999, PhD thesis, Princeton Univ.

\bibitem[Benz \& Paesold(1998)]{1998A&A...329...61B}
Benz, A.~O., \& Paesold, G.\ 1998, A\&A, 329, 61 

\bibitem{bhat}
Bhat, N.D.R. \etal\  2007, ApJ, 665, 618

\bibitem[Blandford(1977)]{1977MNRAS.181..489B} 
Blandford, R.~D.\ 1977, MNRAS, 181, 489 

\bibitem[Burns et al.(2012)]{2012AdSpR..49..433B} Burns, J.~O., Lazio, J., 
Bale, S., et al.\ 2012, Advances in Space Research, 49, 433 

\bibitem{cairns99}
Cairns, I. H., \& Robinson, P. A. 1999, PRL 82, 3066

\bibitem{cam}
Cameron, P. B. \etal\  2005, Nature, 434, 1112

\bibitem{carr83}
Carr, T. D., Desch, M. D., \& Alexander, J. K. 1983, in Physics of the Jovian Magnetosphere, ed. Dessler, A. J. (New York: Cambridge University Press), 226-284

\bibitem{clark12}
Clark, M. A., LaPlante, P. C., Greenhill L. J. 2012, Intl. J. High Perf. Comptuing, in press.


\bibitem{clarke}
Clarke, T.\ 2009, `` Scientific Requirements for the Long Wavelength Array'', Memo 159, LWA Memo Series [http://www.ece.vt.edu/swe/lwa/]




\bibitem[Cordes \& McLaughlin(2003)]{2003ApJ...596.1142C} 
Cordes, J.~M., \& McLaughlin, M.~A.\ 2003, ApJ, 596, 1142 

\bibitem{LWA161}
Craig, J. 2009 ``Long Wavelength Array Station Architecture,'' Ver. 2.0, Memo 161, LWA Memo Series [http://www.ece.vt.edu/swe/lwa/]

\bibitem{Deshpande09}
Deshpande, K.\ B. 2009 {\it A Dedicated Search for Low Frequency Radio Transient Astrophysical Events using ETA}, M.S. Thesis,  Virginia Polytechnic Institute and State University.

\bibitem{des}
Dessenne, C.A.-C. \etal\  1996, MNRAS, 281, 977

\bibitem{dow12}
Dowell, J., Wood, D., Stovall, K., Ray, P.S., Clarke, T., \& Taylor, G.B.
2012, JAI, submitted

\bibitem{Eilek05}
Eilek, J., Hankins, T. \& Jessner 2005, ASP Conf. Ser. 345, 499



\bibitem{elling09}
Ellingson, S.W., Clarke, T.E., Cohen, A. Craig,  J., Kassim,  N.E., Pihlstrom, Y., Rickard, L. J. \& Taylor, G.B. 2009, ``The Long Wavelength Array,'' Proc. IEEE, Vol. 97, No. 8, pp. 1421-1430

\bibitem{Ellingson11}
Ellingson, S.W. 2011, ``Sensitivity of Antenna Arrays for Long-Wavelength Radio Astronomy,'' {\it IEEE Trans.\ Ant.\ \& Prop.\,} 59, 1855, astro-ph/1005.4232.

\bibitem{Eric95}
Erickson, W.C., McConnell, D., \& Anantharamaiah, K.R., 1995, 454, 125

\bibitem[Furlanetto et al.(2006)]{2006PhR...433..181F} Furlanetto, S.~R., 
Oh, S.~P., \& Briggs, F.~H.\ 2006, PhR, 433, 181 

\bibitem{griess}
Griessmeier, J., Zarka, P., \& Spreeuw, H. 2007, A\&A, 475, 359

\bibitem{tim}
Hankins, T. 1973, ApJ, 181, L49

\bibitem[Hansen \& Lyutikov(2001)]{2001MNRAS.322..695H} 
Hansen, B.~M.~S., \& Lyutikov, M.\ 2001, MNRAS, 322, 695 


\bibitem[Helmboldt 
\& Kassim(2009)]{2009AJ....138..838H} Helmboldt, J.~F., \& Kassim, N.~E.\ 2009, AJ, 138, 838 

\bibitem[Hewish]{Hewish} Hewish, A., Bell-Burnell, J., Pilkington, J. D. H., Scott, P. F., \& Collins, R. A. 1968, Nature, 217, 709

\bibitem{How10}
Howard, A.W., Marcy, G.W., Johnson, J.A., Fischer, D.A., Wright, J.T., Isaacson, H., Valenti, J.A., Anderson, J., Lin, D.N.C., \& Ida, S. 2010, Science, 330, 653

\bibitem{Hyman05}
Hyman, S. D., Lazio, T. J. W., Kassim, N. E., Ray, P. S., Markwardt, C. B., \&
Yusef-Zadeh, F. 2005, Nature, 434, 50

\bibitem{Jacoby}
Jacoby, B.A., Lane., W.M., \& Lazio, T.J.W. 2007 ``Simulated LWA-1 Pulsar Observations'' Memo 104, LWA Memo Series [http://www.ece.vt.edu/swe/lwa/]

\bibitem{kant01}
Kantharia, N.G., \& Anantharamaiah, K.R., 2001, JApA, 22, 51

\bibitem{kas93}
Kassim, N.E., Perley, R.A., Erickson, W.C., Dwarakanath, K.S. 1993, 106, 2218

\bibitem{kas98}
Kassim, N.E., \& Erickson, W.C.  1998, SPIE, 357, 740

\bibitem{kass03}
Kassim, N.E., Lazio, T.J.W., Nord, M., Hyman, S.D., Brogan, C.L., LaRosa, T.N., Duric, N. 2003, in Proceedings of the Galactic Center Workshop 2002 - The central 300 parsecs of the Milky Way, Astronomische Nachrichten, Volume 324, Issue S1, p. 73-78.


\bibitem{kas05}
Kassim, N., Perez, M., Junor, W., Henning, P. (eds.), ``From Clark Lake to the
Long Wavelength Array: Bill Erickson's Radio Science'', ASP Conference Series,
Vol., 345, 2005 (Astronomical Society of the Pacific, San Francisco).

\bibitem{kas06}
Kassim \etal\ 2006, ``LWA1+ Scientific Requirements'' Memo 70,
LWA Memo Series [http://www.ece.vt.edu/swe/lwa/]

\bibitem{Kassim+07}
Kassim, N.E.\ \etal\  2007a, ``The 74 MHz System on the Very Large Array,'' {\em Astrophys.\ J.\ Supp.\ Ser.}, Vol.~172, pp.~686--719.


\bibitem{kass10}
Kassim, N.E. \etal\ 2010, ``The Long Wavelength Array (LWA): A
Large HF/VHF Array for Solar Physics, Ionospheric Science, and Solar Radar'',
Proc. Advanced Maui Optical and Space Surveillance Technologies Conf., Sept
14-17, 2010, Maui, Hawaii. Also available as LWA Memo 172.

\bibitem{Kavic+08}
Kavic, M., Simonetti, J.\ H.\ Cutchin, S.\ E., Ellingson, S.\ W.\ \& Patterson, C.\ D.\ 2008 ``Transient Pulses from Exploding Primordial Black Holes as a Signature of an Extra Dimension,'' {\it J. Cosmology \& Astroparticle Physics}, No. 11, 017  http://stacks.iop.org/1475-7516/2008/i=11/a=017.


\bibitem{Lazio09}
Lazio, J., Bastian, T., Bryden, G., Farrell, W. M., Griessmeier, J., Hallinan, G., Kasper, J., Kuiper, T., Lecacheux, A., Majid, W., Osten, R., Shklonik, E., Stevens, I., Winterhalter, D., \& Zarka, P. 2009, Astro2010 whitepaper (arXiv:0903.0873)

\bibitem{Lazio10}
Lazio, T.J.W. {\it \etal\ } (2010), ``Surveying the Dynamic Radio Sky with the Long Wavelength Demonstrator Array,'' AJ, 140, 1995

\bibitem{mad97}
Madau, P., Meiksin, A.,\& Rees, M. J. 1997, ApJ, 475, 429

\bibitem[Malofeev et 
al.(1994)]{1994A&A...285..201M} Malofeev, V.~M., Gil, J.~A., Jessner, A., et al.\ 1994, \aap, 285, 201 

\bibitem[Manchester \etal\ (1973)]{1973ApJ...179L...7M} 
Manchester, R.~N., Taylor, J.~H., \& Huguenin, G.~R.\ 1973, ApJL, 179, L7 




\bibitem{mclaughlin}
McLaughlin, M. A., \etal\  2006, Nature, 439, 817

\bibitem{obenberger11}
Obenberger, K. \& Dowell, J. 2011 ``LWA1 RFI Survey'' Memo 183,
LWA Memo Series [http://www.ece.vt.edu/swe/lwa/]

\bibitem[de Oliveira-Costa \etal\ (2008)]{decosta} de 
Oliveira-Costa, A., Tegmark, M., Gaensler, B.~M., Jonas, J., Landecker, 
T.~L., \& Reich, P.\ 2008, MNRAS, 388, 247 


\bibitem{perley}
Perley, R.A. \& Erickson, W.C.\ 1984, ``A Proposal for a Large, Low Frequency Array Located at the VLA Site'',
Memo 1, LWA Memo Series [http://www.ece.vt.edu/swe/lwa/]

\bibitem{wendy}
Peters, W. M., Lazio, T. J. W., Clarke, T. E., Erickson, W. C., Kassim, N.E. 2011, A\&A, 525, 128

\bibitem{popov}
Popov, M.V. \etal\  2006, {\it Astron. Rep.}, 50, 562

\bibitem{pri08}
Pritchard, J. R., \& Loeb, A. 2008, {\it Phys. Rev. D}, 78, 103511 

\bibitem{pri10}
Pritchard, J. R., \& Loeb, A. 2010, {\it Phys. Rev. D}, 82, 023006

\bibitem[Ransom(2001)]{2001PhDT.......123R} Ransom, S.~M.\ 2001, 
Ph.D.~Thesis

\bibitem[Rees(1977)]{1977Natur.266..333R} 
Rees, M.~J.\ 1977, Nature, 266, 333 

\bibitem{rickard}
Rickard L.J. \etal\ 2010 `` The Long Wavelength Array (LWA): A Large HF/VHF Array for Solar Physics, Ionospheric Science, and Solar Radar'', Ground-Based Instrument Paper for the 2010 NRC Decadal Survey of Solar and Space Physics, Memo 173, LWA Memo Series [http://www.ece.vt.edu/swe/lwa/]

\bibitem{Soriano11}
Soriano, M., R.\ Navarro, L.\ D'Addario, E.\ Sigman \& D.\ Wang (2011), ``Implementation of a Digital Signal Processing Subsystem for a Long Wavelength Array Station,'' {\it Proc. 2011 IEEE Aerospace Conf, Big Sky, MT}.  Also available as LWA Memo~179.

\bibitem{Step07}
Stepkin, S.V. \etal\ 2007, MNRAS, 374. 852

\bibitem{stap11}
Stappers, B.W.\ {\it \etal\ } 2011, \aap, 530, A80 



\bibitem[Ulyanov \etal\ (2006)]{2006IAUJD...2E..12U} 
Ulyanov, O.~M., Zakharenko, V.~V., Konovalenko, A.~A., Lecacheux, A., Rosolen, C., 
\& Rucker, H.~O.\ 2006, IAU Joint Discussion, 2, 

\bibitem[Vachaspati(2008)]{2008PhRvL.101n1301V} 
Vachaspati, T.\ 2008, Physical Review Letters, 101, 141301 

\bibitem{deVos09}
de Vos, M., A.W. Gunst \& R. Nijboer 2009, ``The LOFAR Telescope: System Architecture and Signal Processing,'' {\it Proc. IEEE}, 97, 1431

\bibitem{walm82}
Walmsley, M., \& Watson, W.D., 1982, ApJ, 260, 317 

\bibitem{white05} White, S. 2005, in ``From Clark Lake to the Long
  Wavelength Array: Bill Erickson's Radio Science'', Kassim, N.,
  Perez, M., Junor, W., Henning, P. (eds.), ASP Conference Series,
  176, 345 (Astronomical Society of the Pacific, San Francisco)

\bibitem{wijn11}
Wijnholds, S.J. \& W, A. van Cappellen 2011, {\it IEEE Trans. Ant. \& Prop.}, 59, 1981

\bibitem{york07}
York, J. \etal\ 2007 ``The LWDA Array: An Overview of the Design,
Layout and Initial Results'' Memo 93,
LWA Memo Series [http://www.ece.vt.edu/swe/lwa/]

\bibitem{zarka}
Zarka, P. 2007, Planet. Space Sci., 55, 598
 
\end{thebibliography}
\end{document}